\documentclass[aps,prb,twocolumn,preprintnumbers,amsmath,amssymb]{revtex4}

\usepackage{amsmath}
\usepackage{amsfonts}
\usepackage{graphicx}
\usepackage{mathrsfs}
\usepackage{amsthm}
\usepackage{amssymb}
\usepackage{amsfonts}
\usepackage{mathrsfs}
\usepackage{enumerate}
\usepackage{multirow}
\usepackage[normalem]{ulem}

 \newcommand{\abs}[1]{\left\vert#1\right\vert}
 \newcommand{\set}[1]{\left\{#1\right\}}
 
 \newcommand{\bra}[1]{\big<#1\big\vert}
 \newcommand{\ket}[1]{\big\vert#1\big>}
 \newcommand{\braket}[2]{\big<#1\vert#2\big>}
 \newcommand{\qt}{\tilde{\mathbf{q}}}

 \newcommand{\HUP}{\mathcal{H}^{Uph}}
 
 \newcommand{\Hrelone}{\mathcal{H}^\ast}
 \newcommand{\Hrelonesym}[1]{\mathcal{H}^\ast_{#1}}
 \newcommand{\Hrelonesymstrong}[2]{\mathcal{H}^\ast_{#1,#2}}
 \newcommand{\hqdp}{H^{QDP}}
 \newcommand{\hup}{H^{Uph}}
 \newcommand{\hfr}{H^{int}}
 \newcommand{\hfree}{H^0}

 \newcommand{\q}{\mathbf{q}}
 \newcommand{\x}{\mathbf{x}}
 
 \newcommand{\B}{\mathcal{B}^\ast}
 \newcommand{\Bsym}[1]{\mathcal{B}^\ast_{#1}}
 \newcommand{\Psym}[1]{\mathcal{P}_{#1}}
 \newcommand{\HQSP}{\mathcal{H}^{QSP}}
 \newcommand{\HQSPkph}[1]{\mathcal{H}^{QSP}_{#1}}
 \newcommand{\HQSPxkph}[1]{\tilde{\mathcal{H}}^{QSP}_{#1}}
 \newcommand{\HQSPxxkph}[1]{\tilde{\mathcal{H}}^{QSP\,\perp}_{#1}}
 \newcommand{\hqsp}{H^{QSP}}
 \newcommand{\HQDPkph}[1]{\mathcal{H}^{QDP}_{#1}}
 \newcommand{\HQDPxkph}[1]{\tilde{\mathcal{H}}^{QDP}_{#1}}
 \newcommand{\HQDPxxkph}[1]{\tilde{\mathcal{H}}^{QDP\,\perp}_{#1}}
\begin{document}

\title{Non-orthogonal Theory of Polarons and Application to Pyramidal Quantum Dots}

\author{D. Obreschkow$^{1,3\,}$}
\author{F. Michelini$^{2,3}$}
\author{S. Dalessi$^3$}
\author{E. Kapon$^3$}
\author{M.-A. Dupertuis$^3$}

\affiliation{ $^1\,$Astrophysics, Department of Physics, University of Oxford, Keble
Road, Oxford, OX1 3RH, UK \\
$^2\,$Provence Material and Microelectronics Laboratory (L2MP), 13384 Marseille Cedex 13, France \\
$^3\,$\'{E}cole Polytechnique F\'{e}d\'{e}rale de Lausanne (EPFL), Laboratory of Physics of Nanostructures, CH-1015 Lausanne, Switzerland}
\date{\today}

\begin{abstract}

We present a general theory for semiconductor polarons in the framework of the Fr\"{o}hlich interaction between
electrons and phonons. The latter is investigated using non-commuting phonon creation/annihilation operators associated
with a natural set of non-orthogonal modes. This setting proves effective for mathematical simplification and physical
interpretation and reveals a nested coupling structure of the Fr\"{o}hlich interaction. The theory is non-perturbative
and well adapted for strong electron-phonon coupling, such as found in quantum dot (QD) structures. For those
particular structures we introduce a minimal model that allows the computation and qualitative prediction of the
spectrum and geometry of polarons. The model uses a generic non-orthogonal polaron basis, baptized ``the natural
basis''. Accidental and symmetry-related electronic degeneracies are studied in detail and are shown to generate
unentangled zero-shift polarons, which we consistently eliminate. As a practical example, these developments are
applied to realistic pyramidal GaAs QDs. The energy spectrum and the 3D-geometry of polarons are computed and analyzed,
and prove that realistic pyramidal QDs clearly fall in the regime of strong coupling. Further investigation reveals an
unexpected substructure of ``weakly coupled strong coupling regimes'', a concept originating from overlap
considerations. Using Bennett's entanglement measure, we finally propose a heuristic quantification of the coupling
strength in QDs.

\end{abstract}

\maketitle

\section{Introduction}\label{introduction}

Quantum structures (QSs), such as quantum dots (QDs), are sophisticated solid-state pieces, vital for fundamental
research and novel applications in quantum optics and quantum informatics. Today, QDs find technological use in QD
lasers,\cite{Huffaker 03} infrared photodetectors,\cite{Lee 99} single photon sources~\cite{Santori 01,Pelton 02} or
markers in biology.\cite{Mingyong 01} Cutting-edge research features QDs as medical fluorophores for {\em in vivo
detection} of cell structures such as tumors.\cite{Xingyong 02} Other promising applications for QDs are solar
cells~\cite{Schaller 04} and optical telecommunication.\cite{Klopf 01} The most exciting, yet challenging expectation
relies in the use of QDs as qubit holders and gates for quantum computation.\cite{Loss 98} For fundamental science, QDs
are among the few systems allowing controlled experiments with single energy quanta giving direct access to controlled
quantum entanglement and correlations.

Due to their extreme carrier sensitivity, much interest in QDs relates to carrier relaxation and excitation processes
mediated by various interactions, such as carrier-carrier, carrier-photon and carrier-phonon interactions. As for
carrier-phonon interactions, early perturbative approaches with acoustic phonons resulted in the \emph{bottleneck
concept}.\cite{Bockelmann 90,Benisty 91,Brunner 92,Bockelmann 93} Theses perturbative results predict inefficient
carrier relaxation for a large class of small QDs. Although experimentally verified in certain cases \cite{Urayama
01,Heitz 01}, these predictions failed in many other tests.\cite{Notomi 91,Raymond 95} A definite progress came with
non-perturbative investigations of the deformation potential and Fr\"{o}hlich interaction, revealing the existence of a
strong coupling regime, which is out of reach of perturbative approaches and allows efficient carrier relaxation
through acoustic and optical phonon dynamics respectively.\cite{Inoshita 97,Kral 98,Hameau 99,Verzelen 00} This led to
the new concept of quantum dot polarons (QDPs), which are non-separable fundamental excitations determined by the
carrier-phonon interaction. Within the approximation of monochromatic LO-modes for the Fr\"{o}hlich interaction,
electrons only couple to a finite number of lattice modes as analytically explained through an algebraic decomposition
introduced by Stauber \textit{et al}.\cite{Stauber 00} Their procedure constructs an orthonormalized basis of relevant
lattice modes from the finite set of phonon creation/annihilation operators naturally appearing in the Fr\"{o}hlich
Hamiltonian. This leads to a numerically solvable model of QDPs \cite{Stauber 06}, which can be viewed as an extension
of the work by Ferreira \textit{et al}.\cite{Ferreira 03}

In this work, the polaron problem is tackled from a different viewpoint: the \emph{full} electron-phonon Hamiltonian is
reformulated in terms of non-orthogonal modes, which naturally span all coupled and uncoupled crystal vibrations. The
non-orthogonal structure is preserved from the beginning to the end and exhibits undisputable advantages for
computation and physical understanding. General analytical results applicable to any type of semiconductor QS are
derived in this framework. They are subsequently applied to peculiar pyramidal $C_{3v}$ GaAs QDs, but the same
theoretical scheme could be applied to any other semiconductor QD structure, e.g.~zincblende $InAs$ QDs with $C_{2v}$
symmetry\cite{Gammon 96} or Wurzite $GaN$ QDs with high $C_{3v}$ or $C_{6v}$ symmetry.\cite{Tronc 04}

Section~\ref{theory} considers a general QS populated by an arbitrary number of bound electrons and phonons. We first
introduce a set of non-orthogonal LO-modes, which spans all the LO-modes appearing in the Fr\"{o}hlich interaction.
From there we derive two decoupled subalgebras of non-commuting phonon creation/annihilation operators, which separate
the quantum structure in a subsystem of bound polarons and a subsystem of uncoupled modes (\ref{1st}). The theory
culminates in a non-trivial nested coupling structure of the Fr\"{o}hlich interaction, which has important consequences
when working with any finite number of phonons (\ref{coupling structure}). In section~\ref{QDPs}, we introduce a
minimal non-perturbative model (one electron, one phonon) particularly suitable for the crucial case of QDPs. We
provide an explicit non-orthogonal \emph{polaron} basis, baptized the ``natural basis'' (\ref{sec natural basis}). It
provides a detailed interpretation of the geometries and spectra of low-energy QDPs. We also investigate additional
simplifications resulting from electronic degeneracies (\ref{electronic degeneracies}) and group theoretical
considerations of dot symmetries (\ref{symmetrical qd}). The theory concludes with some key aspects of the
three-dimensional (3D) numerical code (\ref{numerical procedure}), which comprises an adaptive irregular space
discretization for computing the Fr\"{o}hlich matrix elements.

In a second part, the minimal model is applied to realistic pyramidal GaAs QDs with $C_{3v}$ symmetry. \cite{Kapon 04}
Section~\ref{numerics} presents the 3D-geometries of the QDPs and their spectrum, throughout using group theory
(\ref{numerical model},~\ref{results},~\ref{dot size variation}). Explicit comparison with perturbation theory confirms
the existence of a strong coupling regime. Surprisingly, we find significant numerical evidence for a peculiar
substructure inside the strong coupling regime. This leads to the concept of ``weakly coupled strong coupling regimes''
(\ref{substructure}), which can be understood in terms of overlap between confined electrons and coupled modes. Using
Bennett's entanglement measure, we further present a useful alternative characterization of the strong electron-phonon
coupling in QDs (\ref{entanglement}).

In section~\ref{generalQD}, we report on the general polaron properties that could be expected in other QD systems.
Section ~\ref{conclusion} concludes the theory with a short review, and helpful derivations are provided in the
appendix~\ref{appendix}.


\section{Non-orthogonal theory for polaron states}\label{theory}

In this section, we present a theory for polar semiconductor QSs, e.g. dots, wires or wells, in which the carrier
evolution is reasonably described by Fr\"{o}hlich interactions with monochromatic LO-modes. The QSs can contain an
arbitrary number of electrons (within the limitations induced by the Pauli exclusion principle) and an arbitrary number
of phonons. The conservation laws exhibited by the interaction Hamiltonian allow straightforward generalizations to
exciton-polarons or even polarons associated with bigger electron/hole complexes.

\subsection{Polaron Hamiltonian in Quantum Structures}

The model's evolution is dictated by a Hamiltonian composed of a free evolution term and the Fr\"{o}hlich Hamiltonian
$\hfr$,
\begin{equation}
\begin{split}\label{original hamiltonian}
  \lefteqn{H}\qquad      & = \sum_{\mu\,\sigma}\epsilon_\mu a_{\mu\sigma}^\dag a_{\mu\sigma}+\epsilon_{LO}\sum_\q b_\q^\dag b_\q+\hfr \\
  \lefteqn{\hfr}\qquad   & = \sum_{\mu\mu'\sigma\q} M_{\mu\mu'\q} a_{\mu\sigma}^\dag a_{\mu'\sigma} b_\q + h.c. \\
\end{split}
\end{equation}
(Unity operators and tensor products have been omitted.) $a_{\mu\sigma}$, $a_{\mu\sigma}^\dag$ are fermionic
annihilation and creation operators of confined conduction electrons, with $\mu$ labeling an orthogonal set of
stationary wave functions and $\sigma$ being the spin index. The scalars $\epsilon_\mu$ are the free electronic
energies, which are independent of $\sigma$ in the absence of magnetic fields. $b_\q$, $b^\dag_\q$ are the bosonic
annihilation and creation operators of phonons associated with the LO-plane waves $\xi_\q(\x)\equiv\sqrt{
2/V}e^{i\q\cdot\x}$, where $V$ is the quantization volume. $\epsilon_{LO}=\hbar\omega_{LO}$ is the phonon energy
assumed independent of $\q$ (monochromaticity), and $M_{\mu\mu'\q}$ are the Fr\"{o}hlich matrix
elements~\cite{Froehlich 49}
\begin{equation}
\begin{split}\label{froehlich matrix element}
  M_{\mu\mu'\q} = &
  \sqrt{\frac{\hbar\omega_{LO}e^2}{2\varepsilon_0Vq^2}\left(\frac{1}{\varepsilon_\infty}-\frac{1}{\varepsilon_{stat}}\right)}\\
  & \times\ \int_{\mathbb{R}^3}d^3x\ e^{i\q\cdot\x}\psi_{\mu}^\ast(\x)\psi_{\mu'}(\x)\\
\end{split}
\end{equation}
where $\varepsilon_{stat}$ and $\varepsilon_\infty$ are the static and high frequency dielectric constants and
$\psi_{\mu}(\mathbf{x})$ are the (one-particle) electronic wave functions. Since the Hamiltonian (\ref{original
hamiltonian}) is decoupled and symmetrical in spin degrees of freedom, we shall from here on omit the spin indices
$\sigma$.

\subsection{Subsystem of Quantum Structure Polarons}\label{1st}
We shall now apply a non-orthogonal linear transformation to the operator basis $\{b_\q,b^\dag_\q\}$ in order to reveal
two decoupled physical subsystems, the subsystem of ``Quantum Structure Polarons'' (QSPs) and the subsystem of
``Uncoupled Phonons'' (Uphs). This conceptual separation will be reflected in a tensor product decomposition of the
representative Hilbert space.

The matrix elements (\ref{froehlich matrix element}) can be considered as discrete 3-dimensional functions of $\q$.
They obey relations of linear dependance, as can be seen by choosing the electronic wave functions $\psi_{\mu}(x)$ real
(always possible), in which case $M_{\mu\mu'\q}=M_{\mu'\mu\q}$. If there are $N$ orthogonal electron states $\mu$, the
number of such relations is $N(N-1)/2$. The remaining $N(N+1)/2$ matrix elements show no obvious relations of linear
dependance, and we shall temporarily assume that there are the only $N(N-1)/2$ independent relations of linear
dependance. The theory remains valid in the case of additional linear dependencies such as discussed towards the end of
this subsection.

The structure of the Fr\"{o}hlich interaction implies that the number of linearly independent matrix elements
$M_{\mu\mu'\q}$ equals the number of linearly independent lattice modes that appear in the interaction term. This can
be seen explicitly, when reformulating the interaction as
\begin{eqnarray}
  \lefteqn{\hfr}      & \quad\ \ =      & \sum_{\mu\mu'} J_{\mu\mu'} a_\mu^\dag a_{\mu'} B_{\mu\mu'} + h.c. \label{HFr} \\
  \lefteqn{B_{\mu\mu'}} & \quad\ \ \equiv & \frac{1}{J_{\mu\mu'}} \sum_\q M_{\mu\mu'\q} b_\q
  \equiv \sum_\q L_{\mu\mu'\q} b_\q \label{Bop} \\ \nonumber
\end{eqnarray}
where $|J_{\mu\mu'}|^2 = \sum_\q |M_{\mu\mu'\q}|^2$ quantizes the electron-phonon coupling strength, with $J_{\mu\mu'}$
chosen as a positive real. The relations of linear dependance among the matrix elements $M_{\mu\mu'\q}$ trivially
translate to $B^{\dag}_{\mu\mu'}=B^{\dag}_{\mu'\mu}$ and $B_{\mu\mu'}=B_{\mu'\mu}$. The remaining $N(N+1)/2$ linearly
independent phonon operators shall be scanned by a unique pair index $\lambda\equiv\{\mu,\mu'\}=\{\mu',\mu\}$.

The operators $\{B_{\lambda},B^{\dag}_{\lambda}\}$ annihilate and create ``coupled phonons'', that is quanta in terms
of a harmonic oscillator in modes susceptible to interact with electrons via the Fr\"{o}hlich potential. Using
(\ref{Bop}), the wave functions of those modes are given by the inverse Fourier transforms
\begin{equation}
  \Xi_\lambda(\x)=\sum_q L^\ast_{\lambda\q}\xi_q(\x)\equiv\sqrt{\frac{2}{V}}\sum_q e^{i\q\cdot\x}L^\ast_{\lambda\q}\label{phwavefunctions}
\end{equation}
which are manifestly localized in the quantum structure. The set of all modes $\Xi_\lambda(\x)$ is non-orthogonal as
emphasized by the non-diagonal scalar product matrix and the non-diagonal commutator of the corresponding operators
$\{B_{\lambda},B^{\dag}_{\lambda}\}$. Both follow directly from (\ref{phwavefunctions}) and (\ref{Bop}),
\begin{equation}\label{main commutator}
  \Lambda_{\lambda\lambda'}\equiv\left[B_{\lambda},B^\dag_{\lambda'}\right]=
  \big(\Xi_\lambda,{\Xi_{\lambda'}}\big)=\sum_\q
L^\ast_{\lambda'\q}L_{\lambda\q}
\end{equation}
(Round brackets represent the scalar product relative to the quantization volume $V$.) For reasons of physical
interpretation and mathematical simplicity, we skip a possible orthonormalization and preserve the non-orthogonality
for the rest of the theory.

In order to express the full Hamiltonian in terms of the new operators $\{B^{\dag}_\lambda,B_\lambda\}$, we need to
complete them by an operator-set $\{B^{\dag}_\q,B_\q\}$ generating the orthogonal complement $\{\Xi_\q(\x)\}$ of the
coupled modes $\{\Xi_\lambda(\x)\}$. We choose a linear transformation,
\begin{equation}\label{operatorPP}
  B_\q^\dag\equiv\sum_{\q'}{c_{\q\q'}{b_{\q'}}^\dag}\ ,\,\ \ \Xi_\q(\x)\equiv \sum_{\q'}{c_{\q\q'}\xi_\q(\x)}
\end{equation}
A natural and sufficient condition for the coefficients $c_{\q\q'}$ writes $B_\q^\dag\ket{0}=
(\openone_{1\,ph}-\mathcal{P})\,b_\q^\dag\ket{0}$, where $\ket{0}$ is the phonon vacuum, $\openone_{1\,ph}$ the unity
on the subspace of one phonon, and $\mathcal{P}$ is the orthogonal projector on the sub-subspace of coupled one-phonon
states $vect\{B_\lambda^\dag\ket{0}\}$. Thus $(\openone_{1\,ph}-\mathcal{P})$ projects on the one-phonon sub-subspace
of uncoupled modes, and $\{B^{\dag}_\q\}$ creates quanta accordingly called ``uncoupled phonons''. An explicit
derivation of the coefficients $c_{\q\q'}$ is provided in appendix \ref{derivation 1}. From this explicit form it
follows that the modes $\{\Xi_\q(x)\}$ are also mutually non-orthogonal, which again translates to a non-diagonal
commutator of the corresponding creation/annihilation operators $\{B_\q^\dag,B_\q\}$,
\begin{equation}\label{fund com 3}
\left[B_\q,B^{\dag}_{\q'}\right] = \big(\Xi_\q,{\Xi_{\q'}}\big) = \sum_{q''}{c^\ast_{\q\q''}c_{\q'\q''}},
\end{equation}
Indeed, the modes $\{\Xi_\q(x)\}$ constitute an overcomplete set, according to the $N(N+1)/2$ relations of linear
dependence
\begin{equation}
  \sum_\q L^\ast_{\lambda\q}B^\dag_\q = \sum_\q L^\ast_{\lambda\q}\Xi_\q(\x) = 0\ \ \forall\lambda
\end{equation}
However, it is important to note that all coupled modes $\{\Xi_\lambda(\x)\}$ are orthogonal to all uncoupled ones
$\{\Xi_\q(\x)\}$, as emphasized by the following commutators and scalar products
\begin{equation}\label{fund commutator}
  \left[B_\lambda,B^{\dag}_\q\right]=\big(\Xi_\lambda,\Xi_{\q}\big)=0
\end{equation}

The transformations (\ref{Bop}) and (\ref{operatorPP}) constitute a non-orthogonal mapping
$\{b_\q^\dag\}\rightarrow\{B_\lambda^\dag,B_\q^\dag\}$. The inversion
$\{b_\q^\dag\}\leftarrow\{B_\lambda^\dag,B_\q^\dag\}$ is not unique due to the overcompleteness of $\{B_\q^\dag\}$. A
suitable form, consistent with (\ref{Bop}) and (\ref{operatorPP}), is given by
\begin{equation}\label{back transformation}
  b^\dag_\q =
B^\dag_\q+\sum_{\lambda\lambda'}L_{\lambda'\q}(\Lambda^{-1})_{\lambda\lambda'}B^\dag_{\lambda}
\end{equation}
This allows us to express the phonon number operator in terms of the new operators,
\begin{equation}\label{phonon number operator}
  \sum_\q b_\q^\dag b_\q = \sum_{\lambda\lambda'} \big(\Lambda^{-1}\big)_{\lambda\lambda'}B_\lambda^\dag B_{\lambda'}+\sum_\q B_\q^\dag B_\q
\end{equation}

Finally, the full Hamiltonian~(\ref{original hamiltonian}) transforms to
\begin{eqnarray}
  \label{hamiltonian decomposition} \lefteqn{H}     \qquad\ & =      & \hqsp+\hup,\ \hqsp\equiv \hfree+\hfr\qquad \\
  && \nonumber \\
  \label{eq hfree} \lefteqn{\hfree}\qquad\ & \equiv & \sum_\mu\epsilon_\mu a_\mu^\dag a_\mu + \epsilon_{LO}\sum_{\lambda\lambda'} \big(\Lambda^{-1}\big)_{\lambda\lambda'}B_\lambda^\dag B_{\lambda'} \\
  \label{eq hup} \lefteqn{\hup}  \qquad\ & \equiv & \epsilon_{LO}\sum_{\q} B_\q^\dag B_{\q}
\end{eqnarray}
(unity operators and tensor products have been omitted). The fundamental commutators (\ref{fund commutator}) imply the
commutator
\begin{equation}\label{hilbert space decomposition}
  \left[\hqsp,\hup\right]=0
\end{equation}
The latter defines a unique separation in two physical subsystems, expressed by the tensor product decomposition
\begin{equation}\label{hilbert space decomposition}
  \mathcal{H}=\HQSP\otimes\HUP
\end{equation}
such that $\hqsp$ acts trivially in $\HUP$ and $\hup$ acts trivially in $\HQSP$. [(\ref{hilbert space decomposition})
assumes the bosonic symmetrization of the phonon subsystem.] The subsystem represented in $\HQSP$ consists of electrons
and coupled phonons associated with a finite number of $N(N+1)/2$ linearly independent modes $\{\Xi_\lambda(\x)\}$. The
stationary states (i.e.\ eigenstates of $\hqsp$) are likely entangled in electronic and phononic coordinates and will
be referred as to ``quantum structure polarons'' (QSPs). In contrast, the subsystem of ``uncoupled phonons'',
represented in $\HUP$, is a pure phonon-system associated with infinitely many uncoupled bulk modes $\{\Xi_\q(\x)\}$.
Each such mode evolves trivially under the phonon number operator, and thus the quantum structure problem drastically
reduces to solving $\hqsp$ inside $\HQSP$.

This theory remains valid if the Fr\"{o}hlich matrix elements exhibit other linear dependencies than
$M_{\mu\mu'\q}=M_{\mu'\mu\q}$ (and their linear combinations). Indeed, if there are $N'$ linearly independent matrix
elements, $N'<N(N+1)/2$, it suffices to redefine the index $\lambda$ such as to label only the corresponding
independent operators $B_{\mu,\mu'}$. The derivation above stays valid with this redefinition, if any number $N(N+1)/2$
is replaced by $N'$. (e.g. the number of linear dependencies among the uncoupled modes $B_q$ will be reduced to $N'$,
etc.)

It is worth noting that the Hamiltonian $\hqsp$ manifestly conserves the number of electrons, i.e.
\begin{equation}\label{electronconservation}
  \Big[\hqsp,\sum_\mu\epsilon_\mu a_\mu^\dag a_\mu\Big]=0
\end{equation}
This conservation law implies the existence of one coupled mode that only couples to the electron number operator, such
as shown by Stauber et al.\ \cite{Stauber 06} In contrast to their choice, we decide to keep this particular mode in
the system of QSPs. Indeed, even though this mode does not affect the overall electron dynamics, it is located in the
quantum structure and evolves through the creation an reannihilation of intermediate electrons. Therefore, its
stationary solutions are Glauber-coherent states, very different from the stationary phonon-number states of uncoupled
modes.

\subsection{Nested Coupling Structure}\label{coupling structure}
In this section, we shall uncover a nested structure in the Fr\"{o}hlich coupling. This structure implies in particular
that certain states differing by one phonon (e.g.\ a state with one phonon and a state with two phonons) are
exclusively coupled via intermediate higher order states (e.g.\ a state with three phonons). This non-trivial coupling
structure provides some intuition for the form of stationary states and implies a rule to truncate the Hilbert space if
the polaron problem is restricted to a finite number of phonons.

In the following non-perturbative analysis, two states $\ket{\psi}, \ket{\varphi}$ are called ``coupled'' if the
evolution of one state develops a non-vanishing projection on the other, i.e.\ $\bra{\psi}exp(-i\hqsp
t)\ket{\varphi}\neq0$ for at least one $t$. Thus the subspace $\mathcal{H}_a$ coupled to a subspace $\mathcal{H}_b$ is
given by
\begin{equation}\label{coupling rule}
  \mathcal{H}_a=vect\left\{e^{-i\hqsp t}\ket{\varphi}\ : \forall\,t,\ket{\varphi}\in\mathcal{H}_b\right\}
\end{equation}

In order to identify a coupling structure, we first use the conservation of the number of electrons [eq.
(\ref{electronconservation})]. It reveals that coupling structure can be identified individually for each fixed number
of electrons without loss of generality. For the rest of this section, we shall thus restrict our considerations to
some fixed number of electrons $m$ ($m\geq 1$), and take the subspace $\HQSP$ as restricted to $m$ electrons. Second,
we note that $\hqsp$ does not couple orthogonal spin states, and hence the coupling structure can be investigated with
all electrons having the same fixed spin $\sigma$. As $\hqsp$ acts identically on all values of $\sigma$ spins can be
generally neglected (as in the previous section). Third, we use the property that the Fr\"{o}hlich operator $\hfr$
affects phonon numbers by one unit. Hence, it is useful to decompose $\HQSP$ in subspaces associated with different
numbers of phonons $k$,
\begin{equation}\label{HQSP}
\begin{split}
  \HQSP & = \bigoplus_{k=0}^\infty\HQSPkph{k} \\
  \HQSPkph{k} & \equiv vect\Big\{a^\dag_{\nu_1}\cdots a^\dag_{\nu_m}B^\dag_{\mu_1\mu'_1}\cdots
  B^\dag_{\mu_k\mu'_k}\ket{0}\Big\} \\
\end{split}
\end{equation}
The index list $\{\nu_i,\mu_i,\mu'_i\}$ goes over all combinations of electronic indices, such that $\nu_i\neq\nu_j\
\forall\,i\neq j$ (Pauli exclusion principle). Here, $\ket{0}\equiv\ket{0_{electrons}}\otimes\ket{0_{phonons}}$ denotes
the polaron vacuum.

According to the coupling rule (\ref{coupling rule}), the subspace coupled to $\HQSPkph{k}$ is given by
\begin{equation}\label{coupled space}
  vect\left\{e^{-i\hqsp t}\ket{\varphi}\ : \forall\,t,\ket{\varphi}\in\HQSPkph{k}\right\}
\end{equation}

To pinpoint a particularity in the coupling between $\HQSPkph{p}$ and its ``inferior neighbor'' $\HQSPkph{p-1}$, we
shall temporarily restrict the phonon Fock space to at most $p$ phonons ($p>0$). This implicitly requires a truncation
of the Hamiltonian $\hqsp$ equivalent to imposing $B^\dag_\lambda\ket{\varphi}=0\ \forall\
\ket{\varphi}\in\HQSPkph{p}$. We define $\HQSPxkph{p}$ as the sub-subspace of $\HQSPkph{p}$ coupled to $\HQSPkph{p-1}$
within this restriction. Departing from (\ref{coupled space}) with $k=p-1$, $\HQSPxkph{p}$ can be simplified to
(derivation in appendix \ref{derivation 2})
\begin{equation}\label{demo2}
  \HQSPxkph{p}\equiv vect\left\{e^{-i\hfree t}\hfr_+\ket{\varphi}\,: \forall\,t,\ket{\varphi}\in\HQSPkph{p-1}\right\}
\end{equation}
where $\hfree$ is the free evolution (\ref{eq hfree}) and $\hfr_+\equiv\sum_{\mu\mu'} J_{\mu\mu'} a_\mu^\dag
a_{\mu'}B^\dag_{\mu\mu'}$ denotes the phonon creating part of the Fr\"{o}hlich interaction (\ref{HFr}). In physical
terms, (\ref{demo2}) expresses that an electron-phonon state $\ket{\varphi}$, initially containing $p-1$ phonons,
evolves towards a superposition involving a certain $p$-phonon state (by Fr\"{o}hlich interaction). The latter is
generally not an eigenstate of $\hfree$ and its free evolution can span a whole $p$-phonon subspace coupled to the
initial state $\ket{\varphi}$. For further simplification we decompose $\hfr_+\ket{\varphi}$ in eigenstates of
$\hfree$,
\begin{equation}\label{eigenstate decomposition}
  \hfr_+\ket{\varphi}=\sum_\gamma{\mathcal{P}_\gamma\hfr_+\ket{\varphi}}
\end{equation}
where $\gamma$ labels the eigenspaces of $\hfree$ inside $\HQSPkph{p}$, and $\mathcal{P}_\gamma$ are the orthogonal
projectors on all these eigenspaces. As $\mathcal{P}_\gamma$ projects on a $p$-phonon subspace and $\ket{\varphi}$ is a
$(p-1)$-phonon state, we can safely replace $\mathcal{P}_\gamma\hfr_+$ by $\mathcal{P}_\gamma\hfr$, for
$\mathcal{P}_{\gamma}$ annihilates the phonon annihilating part of the interaction $\hfr$. Invoking the relation
$e^{-i\hfree t}\mathcal{P}_\gamma\hfr\ket{\varphi}=e^{-i\epsilon_\gamma t/\hbar}\mathcal{P}_\gamma\hfr\ket{\varphi}$
and substituting (\ref{eigenstate decomposition}) in (\ref{demo2}) gives
\begin{equation}\label{p phonon subsubspace}
  \HQSPxkph{p} = vect\left\{\mathcal{P}_{\gamma}\hfr\ket{\varphi}\,: \forall\,\gamma,\ket{\varphi}\in\HQSPkph{p-1}\right\}
\end{equation}
(since  $e^{-i\epsilon_\gamma t/\hbar}$ for different $\gamma$ are linearly independent functions of $t$.)

The eigenspace projectors $\mathcal{P}_{\gamma}$ act trivially on the subsystem of lattice modes populated by $p$
phonons, since all $p$-phonon states are degenerate (monochromaticity assumption). As for the electron subsystem (here
considered as non-degenerate), the different eigenspaces can be labeled as $\gamma\equiv\{\nu_1,\ldots,\nu_m\}$, where
$\{\ldots\}$ denotes an unordered set and $\nu_i\neq\nu_j\ \forall\,i\neq j$. (Spin indices were omitted according to
the introduction of this section.) The electronic part of the projectors $\mathcal{P}_{\gamma}$ can then be expressed
as
\begin{equation}\label{projector}
  a_{\nu_1}^\dag\cdots a_{\nu_m}^\dag\ket{0_{electrons}}\bra{0_{electrons}}a_{\nu_m}\cdots a_{\nu_1}
\end{equation}
Substituting this expression in (\ref{p phonon subsubspace}), allows to express $\HQSPxkph{p}$ with explicit basis
vectors. After rearrangement and substitution of indices, we find
\begin{equation}\label{subspaces}
  \HQSPxkph{p} = vect\Big\{a^\dag_{\nu_1}\cdots a^\dag_{\nu_m}B^\dag_{\mu_1\mu'_1}\cdots
  B^\dag_{\mu_p\mu'_p}\ket{0}\Big\}
\end{equation}
where $\mu_1=\nu_1$, and $\mu'_1\neq\nu_i\ \forall i=2,\ldots,m$. Expression (\ref{subspaces}) shows that
$\HQSPxkph{p}$ is necessarily a subspace of $\HQSPkph{p}$.

We conclude that if the number of phonons is limited to $p$ ($p\geq1$), the subspace $\HQSPkph{p-1}$ couples to
$\HQSPxkph{p}$, but not to its orthogonal complement $\HQSPxxkph{p}\subset\HQSPkph{p}$.

In conclusion, if for physical or computational reasons the model is truncated to a finite number of phonons $p$, then
the subspace $\HQSPkph{p}$ must be restricted to $\HQSPxkph{p}$. Otherwise non-physical polarons would appear
(contained in $\HQSPxxkph{p}$), that would seem uncoupled and thus unshifted relative to the free spectrum. Such a
precaution was apparently not taken in previous works.\cite{Stauber 06} The particular truncation
$\HQSPkph{p}\rightarrow\HQSPxkph{p}$ also represents an analytical and computational simplification.

\begin{figure}[h]
  \includegraphics{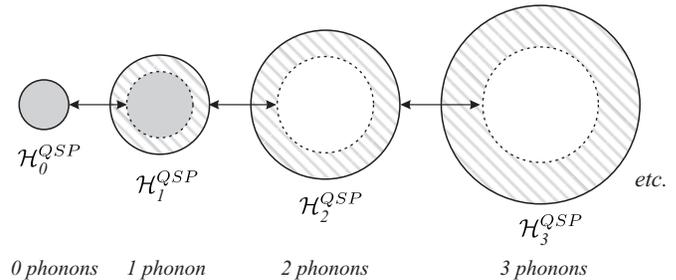}
  \caption{Nested coupling structure exhibited by the Fr\"{o}hlich interaction. Arrows represent couplings that
  do not involve intermediate states with a higher number of phonons (see text). Solid circles contain the subspaces $\HQSPkph{k}$
  for different $k$'s. Dashed circles enclose the subspaces $\HQSPxkph{k}$, whereas
  hatched zones show their orthogonal complements $\HQSPxxkph{k}$ within $\HQSPkph{k}$. The solid gray filling indicates the subspace
  $\Hrelone$ spanned by the natural basis (\ref{natural basis}).}
  \label{fig 01}
\end{figure}

If we release the temporary assumption of a finite phonon number $p$ (or if we take $k<p$), the following statement
holds: $k$-phonon states in $\HQSPxxkph{k}$ do not directly couple to $(k-1)$-phonon states, but can only couple to
$(k-1)$-phonon states via intermediate $(k+1)$-phonon (and higher order) states! In a perturbative approach, these
particular couplings would first appear in the third order of the interaction term. Direct couplings, i.e.~couplings
that do not involve intermediate higher-order states, are represented by the arrows in Fig.\ \ref{fig 01}. This nested
structure provides some insight in the form of stationary polarons (which generally superpose states with different
phonon numbers). For example, stationary superpositions of states from $\HQSPxxkph{k}$ and $\HQSPkph{k-1}$ necessarily
involve a strong contribution of states from $\HQSPxkph{k+1}$. On the other hand, there may be stationary polarons made
of states from $\HQSPxkph{k}$ and $\HQSPkph{k-1}$ with only a minor contribution of states from $\HQSPkph{k+1}$.

\section{One-electron/One-phonon Model of QDPs}\label{QDPs}

In the framework of the general non-orthogonal theory developed above, we shall now propose a minimal non-perturbative
model for polaron states in quantum dots (QDs). The general quantum structure considered so far, is now specified as a
quantum dot: QS$\rightarrow$QD and QSP$\rightarrow$QDP. In such zero-dimensional systems, the monochromaticity
assumption, crucial for the present theory, is fairly precise for the relevant modes (i.e.~wavelengths comparable to
the dot size and thus long compared to the atomic spacing). The model assumes a single electron ($m=1$) populating
different levels while coupling to at most one phonon ($p=1$). It allows to approximate the shifts of the first polaron
levels, which are typically populated at low temperatures, although there may be additional effects arising from
acoustic phonons like dephasing effects.

In the next three subsections, we subsequently investigate QDs with non-degenerate electron levels (\ref{sec natural
basis}), with accidental degeneracies (\ref{electronic degeneracies}), and with symmetry-related degeneracies
(\ref{symmetrical qd}). For each case, we develop a simple non-orthogonal polaron-basis $\B$, baptized the ``natural
basis'', which spans the relevant Hilbert space $\Hrelone$. A similar formalism could be developed for holes (although
acoustic phonons may have to be taken into account there), or for any many-particle complex such as an exciton-, a
trion- or a biexciton-based quantum dot polaron. Stauber and Zimmermann~\cite{Stauber 06} showed that a correction term
must be introduced in the case of non-neutral complexes.

\subsection{Natural Basis}\label{sec natural basis}
We first consider a non-symmetric QD with $N$ non-degenerate electron levels $\mu$. Accordingly there are $N(N+1)/2$
linearly independent coupled modes, spanned by the operators $B^{\dag}_{\mu\mu'}$. By virtue of the coupling structure
developed in section \ref{coupling structure}, the coupled regime of the one-phonon model is properly represented by
the subspace $\Hrelone$ corresponding to gray filling in Fig.\ \ref{fig 01}. It writes
\begin{equation}\label{equ relevant subspace 1}
    \Hrelone\equiv\HQDPkph{0}\oplus\HQDPxkph{1}
\end{equation}
A vector set $\B$, such that $\Hrelone=vect\,\B$ is directly obtained from (\ref{subspaces}),
\begin{equation}\label{natural basis}
  \small\B=\Big\{\ket{\mu;0},B^\dag_{\mu\mu'}\ket{\mu;0}\equiv\sum_q M^\ast_{\mu\mu'\q}\ket{\mu;\q}\
  \forall\,\mu,\mu'\Big\}
\end{equation}\normalsize
where we used the short hands $\ket{\mu;0}\equiv a^\dag_\mu \ket{0}$ and $\ket{\mu;\q}\equiv a^\dag_\mu b^\dag_\q
\ket{0}$, with $\ket{0}=\ket{0_{electrons}}\otimes\ket{0_{phonons}}$ being the polaron vacuum. The vectors in
(\ref{natural basis}) are generally non-orthogonal but linearly independent and $\B$ will be called the ``natural
basis''. All natural basis states are eigenstates of $\hfree$. For each electronic level $\mu$ there is \emph{one}
natural basis state with zero phonons (free energy $\epsilon_\mu$) and there are $N$ natural basis states with one
phonon (free energy $\epsilon_\mu+\epsilon_{LO}$). Since there are $N$ electronic levels $\mu$, the dimension of the
relevant subspace $\Hrelone$ writes
\begin{equation}
  dim(\Hrelone)=card(\B)=N(N+1)
\end{equation}

The requirement to reduce the one-phonon subspace $\HQDPkph{1}$ to $\HQDPxkph{1}$ (section \ref{coupling structure})
reveals the simplifying feature that many product states of electron states and coupled phonons are irrelevant for the
polaron structure (e.g.~$B^\dag_{\mu_2\mu_3}\ket{\mu_1;0} \notin \B$). Therefore, the number of QDPs only scales as
$N^2$ and not as $N^3$ (!), which one might expect from the number $N$ of dot electron states and the number
$\varpropto N^2$ of coupled modes.

\begin{figure}[h]
  \includegraphics{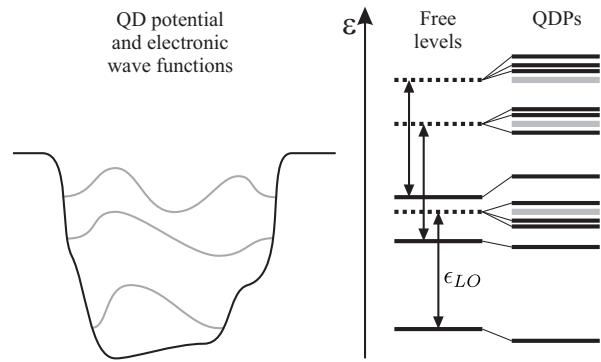}
  \caption{Qualitative structure of the polaron spectrum in the case of three bound, non-degenerate electronic
  levels. Each gray bar indicates 3 additional polaron levels that would result from interactions with two-phonon states.}
  \label{fig 02}
\end{figure}

Fig.~\ref{fig 02} shows the qualitative QDP spectrum in the case of {\em a QD with only three non-degenerate electronic
states}. Gray bars denote additional QDPs that would appear in an extended model including the interaction with
two-phonon states. Those are associated with free states in $\HQDPxxkph{1}$ (i.e. the orthogonal complement of
$\HQDPxkph{1}$ inside $\HQDPkph{1}$). The \emph{connections} between free levels and QDP levels (Fig.~\ref{fig 02}) are
an important outcome of the natural basis. They indicate the free levels from which specific QDPs arise, if one could
gradually introduce the Fr\"{o}hlich interaction. This picture allows a prediction of spectral changes under dot size
variation, since one can generally assume that shifts increase when approaching a resonance of the Fr\"{o}hlich
interaction (\textit{i.e.}\ $\Delta\epsilon=\epsilon_{LO}$).

In conclusion, a complete non-orthogonal polaron basis called the ``natural basis'' (\ref{natural basis}) has been
introduced. It provides a mean for physical understanding of polaron spectra involving low phonon numbers, and
constitutes a simplifying and powerful computational basis (see \ref{numerical procedure}, \ref{numerics}).

\subsection{Accidental Electronic Degeneracies}\label{electronic degeneracies}
This section and the next one point out additional simplifications in the case of electronic degeneracies. In
particular, if a non-degenerate electronic spectrum (e.g.~Fig.~\ref{fig 02}) becomes partially degenerate, for example
by specific dot size adjustment, not only certain QDPs may become degenerate, but some of them will analytically align
with free levels. We shall call such states ``zero-shift polarons'' and show that they are nothing but uncoupled
states, susceptible to become QDPs as soon as the degeneracies are lifted. Thus the relevant Hilbert space $\Hrelone$
can be further reduced, such that spurious zero-shift polarons are  automatically eliminated.

In order to label {\em accidental} degeneracies, the electron index is now expressed as $\mu\equiv(\tau,i)$, where
$\tau=1\ldots n<N$ is an energy index and $i=1,\ldots,g_\tau$ a degeneracy index. The eigenspaces of $\hfree$ inside
the one-phonon subspace $\HQDPkph{1}$ are indexed by $\gamma=\tau$, and the orthogonal projectors
$\mathcal{P}_\gamma\equiv\mathcal{P}_\tau$ on these eigenspaces write
$\mathcal{P}_\tau=\sum_i^{g_\tau}\sum_\q\ket{\tau,i;q}\bra{\tau,i;q}$. Substituting these projectors in (\ref{p phonon
subsubspace}) allows to write the relevant subspace $\Hrelone$ as
\begin{equation}\label{equ relevant subspace 1}
\begin{split}
    & \Hrelone\equiv\HQDPkph{0}\oplus\HQDPxkph{1} \\
    & \text{with}\quad\HQDPxkph{1}=vect\left\{\mathcal{P}_{\tau}\hfr\ket{\mu;0}\ \forall\,\tau,\mu\right\} \\
\end{split}
\end{equation}
Expressing $\HQDPkph{0}$, $\mathcal{P}_\tau$, and $\hfr$ in basis vectors $\ket{\mu;0}$ and $\ket{\mu;\q}$, naturally
provides a basis of $\Hrelone$,
\begin{equation}\label{deg natural basis}
  \small\B=\Big\{\ket{\tau,i;0},\sum_{i'=1}^{g_{\tau'}}\sum_\q M^\ast_{\tau\,i,\tau'\,i';\q}\ket{\tau',i';\q}\
  \forall\,\tau,\tau',i\Big\}
\end{equation}\normalsize
Its dimension is
\begin{equation}\label{dim deg case}
  dim(\Hrelone)=card(\B)=N(n+1)
\end{equation}
where $N$ is the total number of orthogonal electronic states in the dot and $n<N$ is the number of distinct electronic
energies (n=N would be the non-degenerate case.) We note, that even though the number of polarons is smaller in the
degenerate case, the number of modes involved remains the same. Only the number of accessible product states
$\ket{\text{electron}}\otimes\ket{\text{phonon}}$ is reduced. This can be seen from (\ref{deg natural basis}), which,
for a degenerate level $\tau$, yields entangled states similar to
$\sum_{i'}^{g_{\tau'}}B_{(\tau,i)(\tau',i')}^\dag\ket{\tau',i';0}\ \forall\,\tau,i$.

\begin{figure}[h]
  \includegraphics{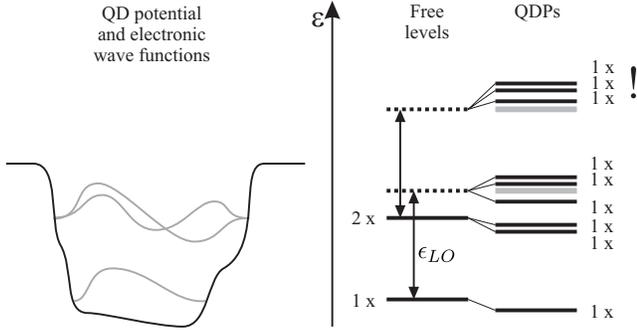}
  \caption{Qualitative structure of the polaron spectrum in the case of two electronic levels, one of which is
  twice degenerate. Each gray bar indicates 3 more polarons that would result from interactions with two-phonon
  states.}
  \label{fig 03}
\end{figure}

As in the previous section, the natural basis (\ref{deg natural basis}) provides a qualitative prediction of the
polaronic spectrum and associates each polaron level with a free level (see example Fig.~\ref{fig 03}). In particular,
we emphasize that the highest free level in the figure only yields 3 orthogonal polaron states and not 6 as one might
expect from pulling together the two uppermost free levels in Fig.~\ref{fig 02}. The particular case of symmetry
related degeneracies is now addressed in the next subsection.

\subsection{Symmetrical Quantum Dots}\label{symmetrical qd}
Additional degeneracies and simplifications may be obtained in the case of QDs invariant under a set of symmetry
operations, generally described by the group of such operations $\mathcal{G}$, i.\ e.\ $[H,g]=0\ \forall
g\in\mathcal{G}$. In such a situation all stationary states satisfy well defined transformation laws, associated with
an irreducible representation (irrep) $\Gamma$ of dimension $d_\Gamma$, which also specifies the respective level
degeneracy. For $d_\Gamma>1$, a degeneracy index $j=1,\ldots,d_\Gamma$, the so-called ``partner function'', labels a
choice of orthogonal states within the same eigenspace. Expressed for passive transformations, the laws read
\begin{equation}\label{transformation law}
  \theta(g)^{-1}\psi_{\Gamma,j} = \sum_{i=1}^{d_\Gamma}\left(D^\Gamma(g)\right)^{-1}_{ij} \psi_{\Gamma,i}\qquad\forall
  g\in\mathcal{G}\quad
\end{equation}
where $D^\Gamma(g)$ is a set of representation matrices that characterize the transformation laws of the partner
function basis, and can be chosen in a suitable way.

Since all stationary states can be associated with a well defined symmetry $(\Gamma,j)$, the Hamiltonian can be
pre-diagonalized by finding an orthogonal decomposition of the Hilbert space in subspaces gathering only states with
symmetry $(\Gamma,j)$. As for the one-phonon/one-electron QD model, this symmetry decomposition writes
\begin{equation}\label{def hrelonesym}
  \Hrelone = \bigoplus_{\Gamma,j}\Hrelonesym{\Gamma,j}\ , \ \ \ \Hrelonesym{\Gamma,j} \equiv \Psym{\Gamma,j}\Hrelone
\end{equation}
where the orthogonal projectors $\Psym{\Gamma,j}$ on the subspaces spanned by
all the states that satisfy the transformation laws~(\ref{transformation law})
for a given symmetry $(\Gamma,j)$ can be written as
\begin{equation}
  \Psym{\Gamma,j}=\frac{d_\Gamma}{\abs{G}}\sum_g \Big(D^\Gamma(g)^{-1}\Big)^\ast_{jj}\,\,\theta^{-1}(g)
\end{equation}

The problem of finding the QDPs reduces to solving $\hqdp$ inside each relevant subspace $\Hrelonesym{\Gamma,j}$
individually. To provide these subspaces with suitable bases, we require a symmetrized eigenstate basis relative to
$\hfree$, \textit{i.e.}~each basis state satisfies the transformation (\ref{transformation law}) for its particular
symmetry $(\Gamma,j)$. Such bases necessarily exist, since $\hfree$ obeys the same symmetry as $H$. To start with we
symmetrize the electron subsystem and the phonon subsystem separately, i.e.
\begin{equation}\label{general sym e-ph basis}
  \begin{split}
  electron:\ \ \set{\ket{\tau,i}}\ & \longleftrightarrow \set{\ket{\Gamma_e,j_e,\alpha_e}}\\
  phonon:\ \set{\ket{0},\ket{\q}}& \longleftrightarrow
  \set{\ket{0},\ket{\Gamma_{ph},j_{ph},\alpha_{ph}}}\quad\\
  \end{split}
\end{equation}
$\alpha_e$ is usually a sequential index with energy, whereas $\alpha_{ph}$
represents a continuous degeneracy index because of the assumption of LO-phonon
monochromaticity. The explicit transformations (\ref{general sym e-ph basis})
can be more subtle than anticipated. An example  will be developed in detail
for the $C_{3v}$ symmetry group in section \ref{numerical model}. These
symmetrized bases allow the construction of a symmetrized basis of the
tensorial products using {\em generalized Clebsch-Gordan coefficients}
$C^{\Gamma,\Gamma_e,\Gamma_{ph}}_{j,j_e,j_{ph}}$ (in the sense of point
groups),
\begin{eqnarray}\label{general sym product basis}
  (a)&&\ket{\Gamma,j;\Gamma_e,\alpha_e;0}=\ket{\Gamma_e=\Gamma,j_e=j,\alpha_e}\otimes\ket{0}\nonumber\qquad\\
     &&\\
  (b)&&\ket{\Gamma,j;\Gamma_e,\alpha_e;\Gamma_{ph},\alpha_{ph}}\nonumber\\
     &&=\sum_{j_e,j_{ph}}C^{\Gamma,\Gamma_e,\Gamma_{ph}}_{j,j_e,j_{ph}}
     \ket{\Gamma_e,j_e,\alpha_e} \otimes\ket{\Gamma_{ph},j_{ph},\alpha_{ph}}\nonumber
\end{eqnarray}
Here $\Gamma$ and $j$ refer to the overall symmetry and $\Gamma_e$ and $\Gamma_{ph}$ satisfy
$\Gamma\subseteq\Gamma_e\otimes\Gamma_{ph}$. The phonon vacuum is always symmetrical, $\Gamma_{ph}=A_1$, and hence the
overall representation of a state with zero phonons will always be identified with the electron representation,
$\Gamma=\Gamma_e$ (eq.\ \ref{general sym product basis}).

A symmetrized expression of the relevant subspace $\Hrelone$ immediately results from equation (\ref{equ relevant
subspace 1}) by replacing the electronic energy index $\tau$ with the pair index $(\Gamma_e,\alpha_e)$. Expressing
$\HQDPkph{0}$ and $\mathcal{P}_{\Gamma_e,\alpha_e,1\,ph}$ in terms of the symmetrized product basis (\ref{general sym
product basis}) directly leads us to a set of non-orthogonal basis vectors, each of which transforms according to
(\ref{transformation law}) for a particular symmetry $(\Gamma,j)$. Those vectors can be regrouped in different
``natural bases'' $\Bsym{\Gamma,j}$ associated with the different subspaces $\Hrelonesym{\Gamma,j}$ defined in
(\ref{def hrelonesym}). The expression of those vectors can be further simplified using the selection rule for the
Fr\"{o}hlich matrix elements, which results directly from the transformation laws and the invariance of the
Hamiltonian,
\begin{equation*}
  \begin{array}{l}
    \bra{\Gamma,j;\Gamma_e,\alpha_e;\Gamma_{ph},\alpha_{ph}}\hfr\ket{\Gamma',j';\Gamma_e'=\Gamma,\alpha_e;0}=0 \\
    \text{unless}\quad(\Gamma,j)=(\Gamma',j')
  \end{array}
\end{equation*}
For the remaining non-vanishing matrix elements, we shall use the notation
\begin{eqnarray}\label{sym matrix element}
  && M^{\Gamma,j}_{(\Gamma_e,\alpha_e,\Gamma_{ph},\alpha_{ph});\alpha_e'} \equiv \\
  && \bra{\Gamma,j;\Gamma_e,\alpha_e;\Gamma_{ph},\alpha_{ph}}\hfr\ket{\Gamma,j;\Gamma,\alpha_e';0} \nonumber
\end{eqnarray}
Finally, the natural bases~$\Bsym{\Gamma,j}$ write
\begin{equation}\label{sym natural basis}
  \Bsym{\Gamma,j}=\left\{
  \begin{array}{l}
    \ket{\Gamma,j;\Gamma,\alpha_e;0},  \\
    \sum_{\Gamma_{ph},\alpha_{ph}} M^{\Gamma,j}_{(\Gamma_e',\alpha_e',\Gamma_{ph},\alpha_{ph});\alpha_e}  \\
    \times\ \ket{\Gamma,j;\Gamma_e',\alpha_e';\Gamma_{ph},\alpha_{ph}}  \\
    \forall\ \alpha_e,\Gamma_e',\alpha_e'
  \end{array}
  \right\}
\end{equation}
The sum goes over all indices, but one assumes that the Clebsch-Gordan coefficients $C$ vanish, when $\Gamma_{ph}$ does
not satisfy a selection rule $\Gamma\subseteq\Gamma_e\otimes\Gamma_{ph}$. These bases are mutually orthogonal, but the
vectors in each individual basis remain non-orthogonal.

The overall symmetry index of an arbitrary natural basis state is always equal to the symmetry index of the involved
pure electron state (zero-phonon state), see equations (\ref{sym matrix element},\ref{sym natural basis}). Hence, if an
existing representation $\Gamma$ is absent in the considered set of bound electrons, there are \emph{no $\Gamma$-like
polaron states}, even though we necessarily have $\Gamma$-like phonon states!

\begin{figure}[h]
  \includegraphics{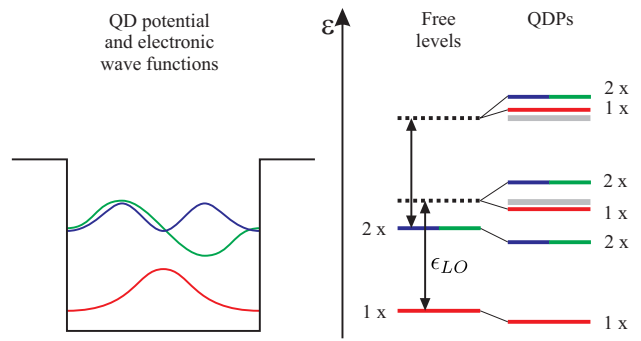}
  \caption{(Color online) Qualitative structure of the polaron spectrum in the case of two electronic levels, one of which is
  twice degenerate due to the dot symmetry. The levels correspond to two different representations $\Gamma$ marked in
  red and blue/green. The latter has two dimensions, characterized by the function $j$. Each gray bar indicates 3 more
  polarons that would result from interactions with two-phonon states.}
  \label{fig 04}
\end{figure}

Like in the previous two sections the natural bases (\ref{sym natural basis})
provide a prediction of the polaronic spectrum. Fig.~\ref{fig 04} shows the
particular case of two bound electronic levels, where the second level is twice
degenerate by virtue of the underlying dot symmetry. In particular, we
emphasize the appearance of degenerate polaron levels, which can be associated
with both, a degenerate \emph{or} non-degenerate electron level.

The dimensionality of the different subspaces $\Hrelonesym{\Gamma,j}$ can be derived from the number of natural basis
states for a given symmetry $(\Gamma,j)$,
\begin{equation}\label{dim sym case}
  dim(\Hrelonesym{\Gamma,j})=card(\Bsym{\Gamma,j})=n^\Gamma(1+n)
\end{equation}
$n^\Gamma$ is the number of {\em distinct} electronic energies with a given symmetry $\Gamma$ ($\alpha_e=1,
\dots,n^\Gamma$), and $n$ denotes the total number of distinct electronic energies ($n=\sum_\Gamma n^\Gamma$). The
dimension (\ref{dim sym case}) is independent of the partner function $j$ in agreement with the feature that those
functions can be defined arbitrarily inside a given representation $\Gamma$. Since
$\sum_{(\Gamma,j)}dim(\Hrelonesym{\Gamma,j})$ equals $dim(\Hrelone)$, given in (\ref{dim deg case}), expression
(\ref{dim sym case}) is a consistent refinement of the full dimension.

\subsection{Computational Aspects}\label{numerical procedure}
Finding the polaron spectrum of the one-electron/one-phonon model reduces to
diagonalizing the Hamiltonian $\hqdp$ inside the low dimensional subspace
spanned by the natural basis $\B$ given in (\ref{natural basis}), (\ref{deg
natural basis}) or (\ref{sym natural basis}), depending on the physical
situation. Although the number of Fr\"{o}hlich matrix elements for the
interaction with LO-phonons has been minimized by the subspace reduction, their
prerequisite computation can be numerically intensive for the arbitrary 3D
wavefunctions that one should consider in a general case (see section
\ref{numerics} where a single wavefunction is typically sampled on $10^6$
points). To alleviate this issue we have developed an original
$\emph{adaptive}$, $\emph{irregular}$ discretization of the reciprocal space
for lattice modes, and shown that it was an efficient method, also applicable
when working directly with a \emph{non-orthogonal} basis.

\begin{figure}[h]
  \includegraphics{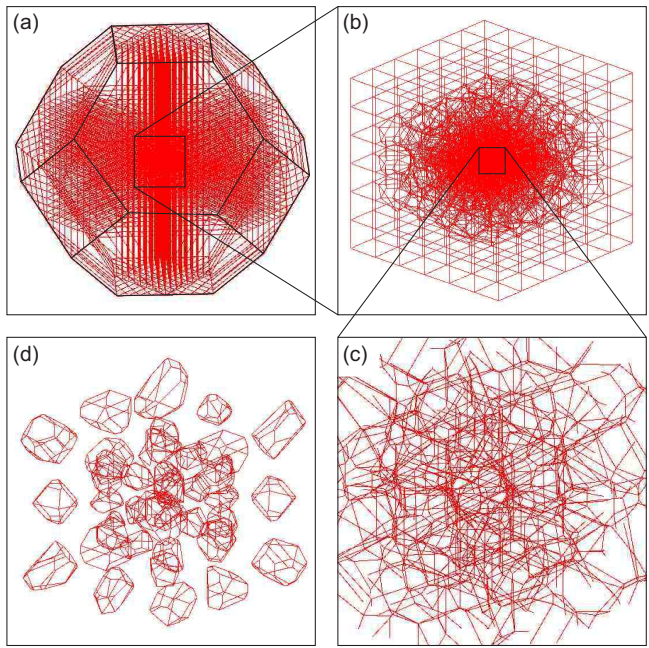}
  \caption{(Color online) Polygonal Wigner-Seitz cells of the irregular $\q$-space discretization.
  (a)-(c) zoom into first Brillouin
  zone of a BCC lattice. (d) A few selected Wigner-Seitz cells from a region
  similar to (c).}
  \label{fig 05}
\end{figure}

The numerical benefit of an irregular reciprocal space discretization relies on the fast variation of the Fr\"{o}hlich
matrix elements in function of the normal mode wave vector $\q$ in certain well localized domains. Increasing the local
point density only in those domains remarkably improves the numerical precision with a minor increase of the required
computational resources. To generate a well adapted irregular $\q$-space discretization, we start with a regular coarse
mesh covering the first Brillouin zone of the underlying lattice. Then the various Fr\"{o}hlich matrix elements are
evaluated for all wave vectors $\q$ of the given mesh. This requires a preliminary computation of the volumes
associated with each mesh node, taken as the volume of the respective Wigner-Seitz cells (Appendix~\ref{froehlich}).
The nearest neighbors with the highest difference between their Fr\"{o}hlich elements are added a new node in between,
which provides the initial mesh for the next iteration. This algorithm is repeated until the maximal difference between
neighboring Fr\"{o}hlich matrix elements falls below a preset threshold. Fig.~\ref{fig 05} shows the Wigner-Seitz cells
generated with this technique in the case of the first Brillouin zone of a body-centered cubic lattice.

After the generation of the irregular $\q$-space discretization and the
computation of the respective Fr\"{o}hlich matrix elements, the set of natural
basis vectors is evaluated, allowing to diagonalize the relevant Hamiltonian
$\hqdp$.

Let us finally evaluate the numerical value of working with a non-orthogonal
basis. The sole consequence is that the standard eigenvalue problem becomes a
\emph{generalized eigenvalue problem}, that is an equation of the type
\begin{eqnarray}
  && \bar{\bar{H}}^{QDP}\ket{\psi}=\epsilon \bar{\bar{S}}\ket{\psi}, \\
  && \bar{\bar{H}}^{QDP}_{\alpha\alpha'}\equiv\bra{\alpha}\hqdp\ket{\alpha'},
  \quad\bar{\bar{S}}_{\alpha\alpha'}\equiv\braket{\alpha}{\alpha'}\neq\delta_{\alpha,\alpha'}\qquad\nonumber
\end{eqnarray}
where $\ket{\alpha}$ are the natural basis states, and
$\bar{\bar{S}}_{\alpha\alpha'}$ is the so-called ''mass matrix''. This
trade-off is advantageous, since optimized packages for the generalized
eigenvalue problem are widely available, and one gets rid of an additional
basis change involving a Gram-Schmidt decomposition (often requiring enhanced
precision for small scalar products). This is a positive numerical byproduct of
the non-orthogonal theory.

With the set of tools presented above, a spectral precision down to 0.01meV for typical QDPs can be reached in
characteristic computation times of a few minutes using a present-day standard processor (3\,GHz, 32\,bit). The most
computer intensive part is the preliminary evaluation of the Fr\"{o}hlich matrix elements.


\section{Application to Pyramidal QDs}\label{numerics}

\subsection{Symmetrical Model and Non-Orthogonal Basis}\label{numerical model}

\begin{figure}[h]
  \includegraphics{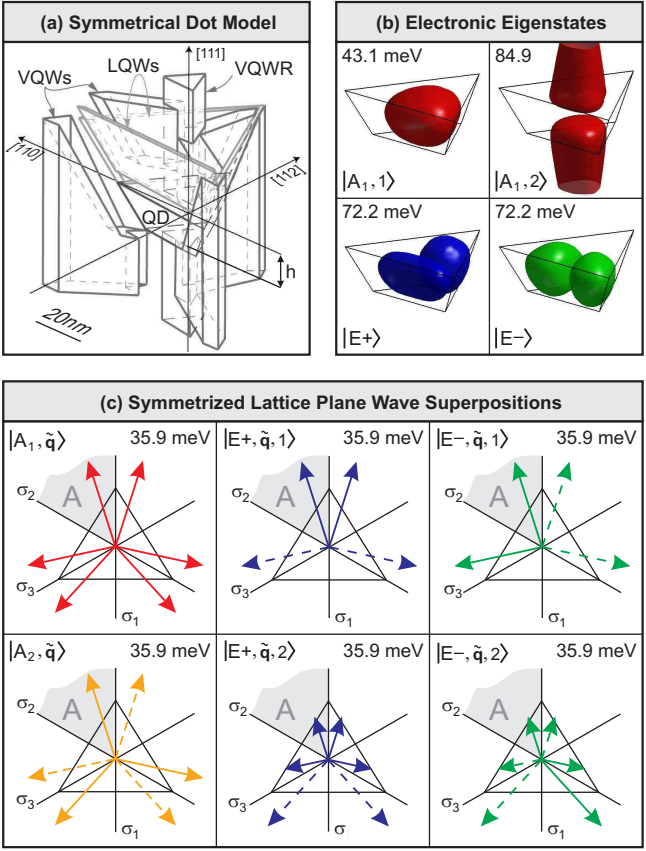}
  \caption{(Color) (a) Numerical model~\cite{Michelini 05} of realistic heterostructure
with pyramidal QD, vertical quantum wire (VQWR), vertical quantum wells (VQWs)
and lateral quantum wells (LQWs). (b) Isosurfaces of envelope functions of the
stationary single electron states. (c) Particular plane wave superpositions of
symmetrized one-phonon basis. The length of the arrows represents the relative
amplitude and dashed arrows have opposite phase.}
  \label{fig 06}
\end{figure}

In this section, we apply the minimal model for the non-orthogonal theory (section \ref{QDPs}) to a realistic pyramidal
GaAs/AlGaAs QD.\cite{Kapon 04} This dot is part of a complex heterostructure represented by the geometrical model shown
in Fig.~\ref{fig 06}a.\cite{Michelini 05} We will take full advantage of the underlying $C_{3v}$-symmetry group, which
exhibits only three irreps $A_1$, $A_2$ and $E$. The latter is two-dimensional and a possible basis results from
symmetrizing $E$-states with respect to the symmetry plane $\sigma_1$ (spanned by the [111] and [112] crystalline
directions in GaAs/AlGaAs). Thereby the partner function $j=\pm$ is identified with the parity index relative to
$\sigma_1$. In graphical representations we shall consistently apply the color scheme: $A_1$ (red), $A_2$ (yellow),
$E+$ (blue), $E-$ (green). QDPs will be computed using the symmetrized natural basis introduced in section
\ref{symmetrical qd}. This basis will be derived analytically in three stages: (1) individual symmetrization of
electronic and phononic eigenstates of $\hfree$, (2) construction of a symmetrized product basis using Clebsch-Gordan
coefficients, and (3) derivation of the symmetrized natural bases for the relevant subspaces $\Hrelonesym{\Gamma,j}$.

First, we shall find symmetrized electron and phonon bases. As for the bound electron, all eigenstates of $\hfree$ are
automatically symmetrized and hence the task reduces to finding these eigenstates. This was recently achieved by
Michelini \textit{et al.}.\cite{Michelini 05} using an effective mass model. For a dot height $h=10nm$, there are two
$A_1$-like levels (non-degenerate) and one $E$-like level (twice degenerate) as shown in Fig.~\ref{fig 06}b. In the
standard notation of section \ref{symmetrical qd}, i.\ e.\ $\ket{\Gamma_e\ j_e,\alpha_e}$, those states write
\begin{equation}
  \Big\{\ket{A_1,1},\ket{E\pm},\ket{A_1,2}\Big\}\ \ electron\ basis
  \label{electron basis}
\end{equation}
where the index $j_e$ has been omitted in the case of the one-dimensional $A_1$-representations and the index
$\alpha_e$ has been omitted for the unique $E$-level. We note that there are no $A_2$-like electron states at low
energy, which immediately predicts that there will be \emph{no} $A_2$-like QDPs (section \ref{symmetrical qd}). For the
phonons (taken as bulk phonons) the symmetrization is inasmuch different as the eigenstates of $\hfree$, such as plane
waves $\xi_\q(\x)$, are \emph{not} automatically symmetrized. This feature relies on the monochromaticity assumption
rendering all normal modes degenerate. A symmetrized eigenstate basis is properly derived in Appendix~\ref{c3v
phonons}. The resulting basis states superpose six (or four) plane waves, such that the directions of the different
wave vectors are mutually related by symmetry operations (see Fig.~\ref{fig 06}c). We shall label such states with the
respective vector $\qt$ belonging to the subset $\mathcal{A}$, which constitutes a sixth of the reciprocal space. In
the case of $E$-like superpositions there are two orthogonal states associated with the same vector $\qt$. They will be
distinguished through the additional index $\chi=1,2$ (discussion in Appendix~\ref{c3v phonons}). Finally the phonon
basis writes
\begin{equation}
  \set{
    \begin{array}{l}
      \ket{0},\ket{A_1,\qt},\ket{A_2,\qt},  \\
      \ket{E\pm,\qt,\chi}
    \end{array}
  }\ \ phonon\ basis
  \label{phonon basis}
\end{equation}
where $\ket{0}$ is the phonon vacuum state (0meV), whilst all
other states are one-phonon states (35.9meV).

Second, we construct a symmetrized product basis from (\ref{electron basis})
and (\ref{phonon basis}) according to Eq.(\ref{general sym product basis}). The
explicit derivations given in Appendix~\ref{c3v tensor product basis} yield
states of the form
\begin{equation}
  \set{
    \begin{array}{l}
      \ket{\Gamma\ j;\Gamma_e,\alpha_e;0},  \\
      \ket{\Gamma\ j;\Gamma_e,\alpha_e;\Gamma_{ph},\qt,\chi}
    \end{array}
  }\ \ product\ basis
  \label{product basis}
\end{equation}
where the first ket represents states with zero phonons and the latter states with one phonon.

Third, we write the natural basis of the relevant Hilbert subspace of
$\Hrelone$ according to the general theory (section \ref{symmetrical qd}). This
basis decomposes in symmetry-subspaces $\Hrelonesym{A_1}$ (8 dimensions),
$\Hrelonesym{E+}$ (4 dimensions), $\Hrelonesym{E-}$ (4 dimensions). Yet, in our
particular case the most energetic natural basis states yield energies above
the first two-phonon state. To remain consistent with the one-phonon
assumption, we shall neglect those states. Thereby the dimensions reduce to 6
($\Hrelonesym{A_1}$), 3 ($\Hrelonesym{E+}$) and 3 ($\Hrelonesym{E-}$). The
respective natural bases result from the general expressions (\ref{sym natural
basis}) and are given in Tab.~\ref{natural basis A1} and Tab.~\ref{natural
basis E}.

\begin{table}[h]
  \begin{center}
    \begin{tabular}{|r|l|c|}
      \hline
        \ meV\ & expressed as symmetrized product states\ & subspace \\
      \hline
        43.1   & $\ket{A_1;A_1,1;0}$ & \multirow{3}{1.1cm}{$\Hrelonesymstrong{A_1}{1}$} \\
      \cline{1-2}
        79.0   & $\sum_{\qt}\ket{A_1;A_1,1;A_1,\qt}\bra{...}\hfr\ket{A_1;A_1,1;0}$ & \\
      \cline{1-2}
        108.1  & $\sum_{\qt,\chi}\ket{A_1;E;E,\qt,\chi}\bra{...}\hfr\ket{A_1;A_1,1;0}$ & \\
      \hline
        84.9   & $\ket{A_1;A_1,2;0}$ & \multirow{3}{1.1cm}{$\Hrelonesymstrong{A_1}{2}$} \\
      \cline{1-2}
        79.0   & $\sum_{\qt}\ket{A_1;A_1,1;A_1,\qt}\bra{...}\hfr\ket{A_1;A_1,2;0}$ & \\
      \cline{1-2}
        108.1  & $\sum_{\qt,\chi}\ket{A_1;E;E,\qt,\chi}\bra{...}\hfr\ket{A_1;A_1,2;0}$ & \\
      \hline
    \end{tabular}
    \caption{Natural basis states of the subspace $\Hrelonesym{A_1}$. The bra
    $\bra{...}$ is the adjoint of the preceding ket. The two subspaces
    $\Hrelonesymstrong{A_1}{1}$ and $\Hrelonesymstrong{A_1}{2}$ are defined by the three basis vectors on their left. They
    constitute so-called ``weakly coupled strong coupling regimes'', discussed in section \ref{substructure}.}
    \label{natural basis A1}
  \end{center}
\end{table}

\begin{table}[h]
  \begin{center}
    \begin{tabular}{|r|l|}
      \hline
        \ meV\ & expressed as symmetrized product states \\
      \hline
        72.2 & $\ket{E\pm;E;0}$ \\
      \hline
        79.0 & $\sum_{\qt,\lambda}\ket{E\pm;A_1,1;E,\qt,\lambda}\bra{...}\hfr\ket{E\pm;E;0}$ \\
      \hline
       108.1 & $\sum_{\Gamma_{ph},\qt,\chi}\ket{E\pm;E;\Gamma_{ph},\qt,\chi}\bra{...}\hfr\ket{E\pm;E;0}$ \\
      \hline
    \end{tabular}
    \caption{Natural basis states of subspaces $\Hrelonesym{E+}$ and $\Hrelonesym{E-}$. The bra
    $\bra{...}$ is the adjoint of the preceding ket.}
    \label{natural basis E}
  \end{center}
\end{table}

\subsection{Stationary States and Strong Coupling}\label{results}

\begin{figure}[h]
  \includegraphics{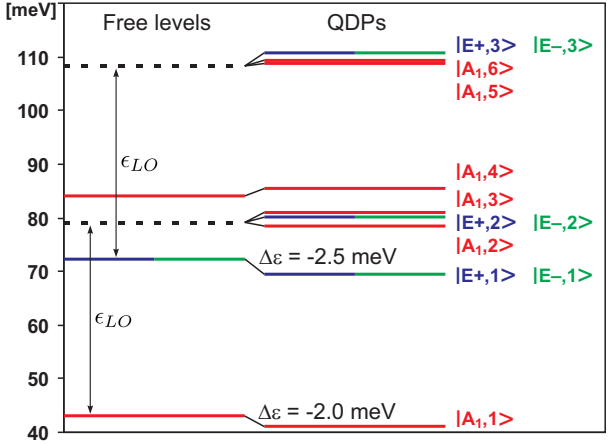}
  \caption{(Color) Spectrum of low energy quantum dot polarons (QDPs). Inside each symmetry subspace the states have been labeled with
  an energy index $m$, such that $m=1$ corresponds to lowest energy state of a given symmetry.}
  \label{fig 07}
\end{figure}

\begin{figure}[h]
  \includegraphics{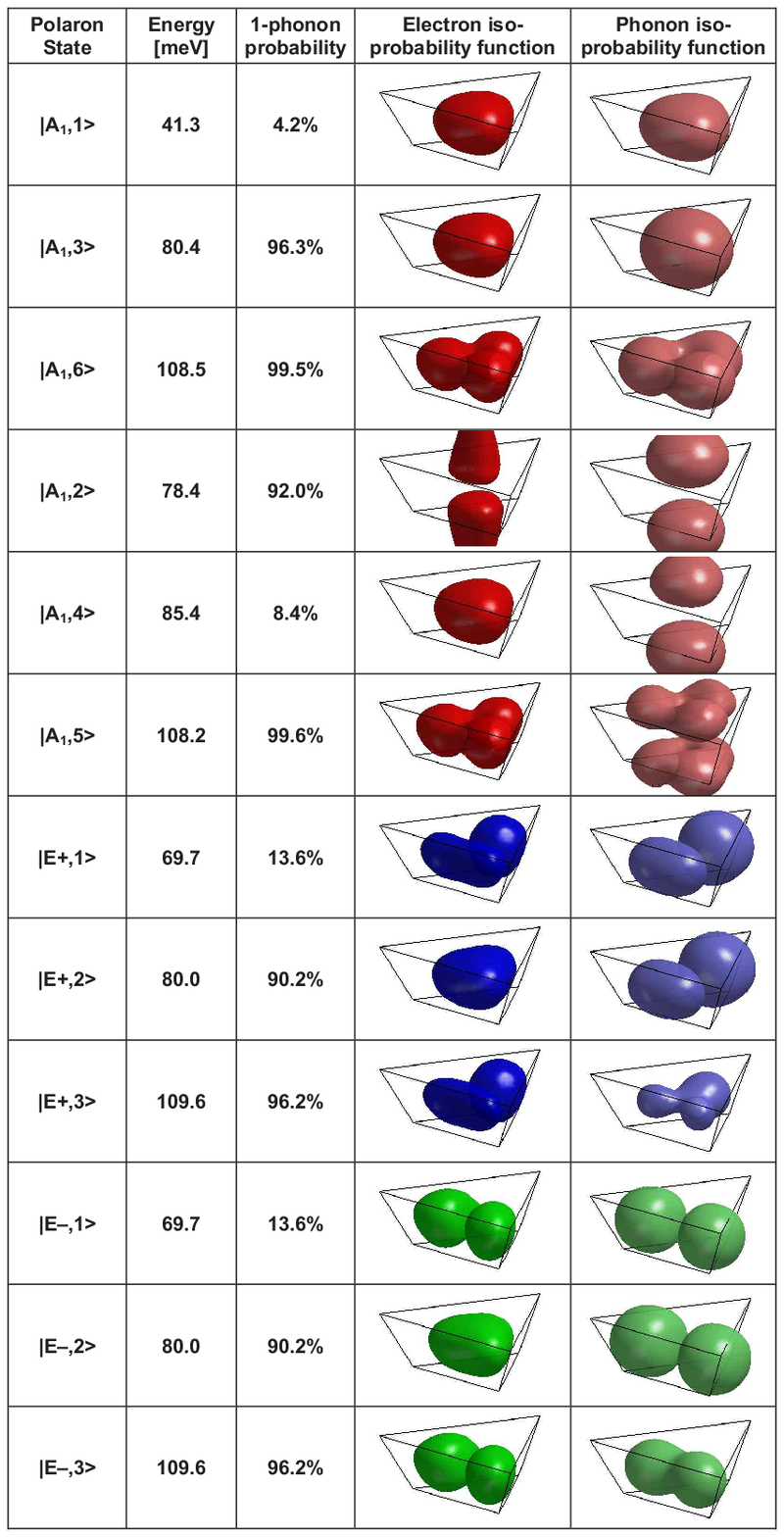}
  \caption{(Color) Geometrical representation of QDPs. The two right columns show isosurfaces of the electronic and vibrational
  probability density functions in direct space.}
  \label{fig 08}
\end{figure}

The problem of finding the stationary dot states, \textit{i.e.}~QDPs, consists in the
eigenvalue problem
\begin{equation}
  \hqdp\ket{\Gamma\,j,m} = \epsilon_{\Gamma,m}\ket{\Gamma\,j,m}
\end{equation}
where $m$ is a sequential energy index inside a particular symmetry
$(\Gamma,j)$. This eigenvalue equation was solved individually inside each of
the three decoupled subspaces $\Hrelonesym{A_1}$, $\Hrelonesym{E+}$ and
$\Hrelonesym{E-}$ using the enhanced matrix diagonalization method outlined in
section \ref{numerical procedure}. The three resulting spectra are given in
Fig.~\ref{fig 07} (red, green, blue). A geometrical representation of the
corresponding polaron states is shown in Fig.~\ref{fig 08}, where the closed
surfaces are isosurfaces of the electronic and vibrational probability density
functions. Those functions were obtained by computing the respective partial
traces,
\begin{eqnarray}
  \rho_{lattice}(\x)  & = & \bra{\x}Tr_{electron}\Big(\ket{\Gamma\,j,m}\bra{\Gamma\,j,m}\Big)\ket{\x}\qquad \\
  \rho_{electron}(\x) & = & \bra{\x}Tr_{lattice}\Big(\ket{\Gamma\,j,m}\bra{\Gamma\,j,m}\Big)\ket{\x}
\end{eqnarray}

The two-dimensional representation $E$ necessarily exhibits a spectrum
consisting of twice degenerate levels, each of which is associated with one
state in $\Hrelonesym{E+}$ and one state in $\Hrelonesym{E-}$. Each
superposition $c_+\ket{E+,m}+c_-\ket{E-,m}$ is again a stationary state.

Both the ground level and the first excited level yield negative energy
shifts. This is consistent with the general feature that the ground level of
\emph{each representation} is necessarily lowered with respect to corresponding
free level. The numerical values of these shifts are $\Delta\varepsilon=-2.0\
meV$ and $\Delta\varepsilon=-2.5\ meV$. The same shifts computed with 2nd order
perturbation theory are $\Delta\varepsilon=-7.0\ meV$ and
$\Delta\varepsilon=-17.4\ meV$, respectively. This manifest large failure of a
perturbative approach clearly confirms the existence of a strong coupling
inside \emph{both} irreps ($A_1$, $E$).

\subsection{Coupling Substructure}\label{substructure}

For further characterization of the coupling regime it is interesting to
consider the two state sets $\mathcal{S}_1$ and $\mathcal{S}_2$, defined as
\begin{equation}
\begin{split}
  \mathcal{S}_1 & \equiv \set{\ket{A_1,1},\ket{A_1,3},\ket{A_1,6}} \\
  \mathcal{S}_2 & \equiv \set{\ket{A_1,2},\ket{A_1,4},\ket{A_1,5}} \\
\end{split}
\end{equation}
The $A_1$-states of Fig.~\ref{fig 08} have been ordered according to these sets. We demonstrated numerically that
states in $\mathcal{S}_1$ are to a good approximation contained in the subspace
$\Hrelonesymstrong{A_1}{1}\subset\Hrelonesym{A_1}$ defined in Tab.~\ref{natural basis A1}. Indeed, the norms of their
projections on $\Hrelonesymstrong{A_1}{1}$ exceed 95\% of the full norms. With the same accuracy the states in
$\mathcal{S}_2$ are contained in $\Hrelonesymstrong{A_1}{2}\subset\Hrelonesym{A_1}$. In other words, the two subspaces
$\Hrelonesymstrong{A_1}{1}$ and $\Hrelonesymstrong{A_1}{2}$ appear {\em reasonably decoupled}, although the whole
subspace $\Hrelonesym{A_1}\equiv\Hrelonesymstrong{A_1}{1}\oplus\Hrelonesymstrong{A_1}{2}$ constitutes a strong coupling
regime. Therefore the strong coupling regime must reside inside the two subspaces $\Hrelonesymstrong{A_1}{1}$ and
$\Hrelonesymstrong{A_1}{2}$ individually, and they may be referred to as ``weakly coupled strong coupling regimes''.
The physical reason for this particular structure relies in the geometry of the vibrational density function
$\rho_{lattice}(\x)$. Fig.~\ref{fig 08} shows that states in $\mathcal{S}_1$ have a vibrational component, which is
vertically centered in the dot, whereas the states in $\mathcal{S}_2$ have \emph{two} centers of vibration splitting
the isosurface in two parts. Indeed the subspace $\Hrelonesymstrong{A_1}{1}$ is spanned by two one-phonon states with
vertically centered vibrational density and one zero-phonon state with centered electronic density. The resulting
overlap leads to a strong interaction between electrons and phonons. The same conclusion applies to the subspace
$\Hrelonesymstrong{A_1}{2}$, where the density functions are vertically split in two parts. One the other hand, this
picture reveals that the mutual overlap between $\Hrelonesymstrong{A_1}{1}$ and $\Hrelonesymstrong{A_1}{2}$ is
considerably smaller.

The concept of weakly coupled subspaces $\Hrelonesymstrong{A_1}{1}$ and
$\Hrelonesymstrong{A_1}{2}$ provides a direct tool for interpretation of the
spectrum in Fig.~\ref{fig 07}. In particular, the ground levels of each
subspace, \textit{i.e.}~$\ket{A_1,1}$ \emph{and} $\ket{A_1,2}$, are necessarily lowered
relative to the corresponding free levels. Analogically, the most excited
levels of each subspaces, \textit{i.e.}~$\ket{A_1,5}$ and $\ket{A_1,6}$, are both
raised. Their mutual splitting remains very small as they are to a good
approximation uncoupled.

Finally, we emphasize that the novel concept of ``weakly coupled strong coupling regimes'', represented by the
subspaces $\Hrelonesymstrong{\Gamma\,j}{1}$ and $\Hrelonesymstrong{\Gamma\,j}{2}$,$\ldots$, is very general and
potentially applicable to all QDs. If the matrix element integral is close to zero due to the mutual orthogonality of
the electronic wave functions, these subspaces can be treated as decoupled in a good approximation. This idea is
straightforward when working with the natural basis, and thus represents a further advantage of using non-orthogonal
basis states.

\subsection{Entanglement measure and strong coupling, decoherence and relaxation}\label{entanglement}

An alternative characterization of the strong coupling is reflected in the
entanglement of stationary dot states, \textit{i.e.}~QDPs. We have computed the
entanglement between electronic and and phononic coordinates using the standard
measure introduced by Bennett \textit{et al.}.\cite{Bennett 96} For pure states,
\begin{equation}\label{def entanglement}
  Ent\big(\ket{\Gamma\,j,m}\big)\equiv-\sum_{i=1}^N{\abs{c_i}^2log_N\abs{c_i}^2}
\end{equation}
where the coefficients $c_i$ are given by the diagonal Schmidt decomposition
\begin{equation}\label{schmidt form}
\ket{\Gamma\,j,m} = \sum_{i=1}^N c_i\ket{e_i}\otimes\ket{ph_i}
\end{equation}
It is known that this form always exists for every particular ket $\ket{\Gamma\,j,m}$, but the computation of the
$\ket{\Gamma\,j,m}$-dependent orthogonal vectors  $\set{\ket{e_i}}$ and $\set{\ket{ph_i}}$, in the electronic and
phononic Hilbert spaces respectively, {\em now requires a preliminary orthogonalization of the quantum dot phonon
basis} $B^\dag_{\mu\mu'}\ket{0}$ which can be elegantly performed via a Choleski decomposition of the $\Lambda$ matrix.
A subsequent singular value decomposition (SVD) of the coefficient matrix in the tensorial product basis will deliver
$c_i$. It is remarkable that the sum over $i$ in~(\ref{schmidt form}) can be limited to the number of electron states
$N$ because of the properties of the SVD. The entanglement measure~(\ref{def entanglement}) can vary between 0
(non-entangled) and 1 (fully entangled), and is also often equivalently called the ''entropy of mixing'' of the two
subsystems in state $\ket{\Gamma\,j,m}$.

Tab.~\ref{tab entanglement} shows the entanglement of the QDPs presented in Fig.~\ref{fig 08}. Weakly entangled states
($Ent\lesssim 0.1$) are nearly simple product states of electrons and phonons. In the present case, such a picture
applies to the two QDPs $\ket{A_1,1}$ and $\ket{A_1,3}$. Numerically, they consist to 99.5\% of two natural basis
states, which \emph{both} involve the \emph{same electronic state} $\ket{A_1,1}$ (first two basis states in
Tab.~\ref{natural basis A1}). All other QDPs are strongly entangled ($Ent\gtrsim 0.1$) with no adequate perturbative
picture. Particularly strong entanglement is found in the states $\ket{A_1,5}$ and $\ket{A_1,6}$, which involve
peculiar Bell-state superpositions inside the $E$-representation of the form
$\ket{E+}\otimes\ket{E+}+\ket{E-}\otimes\ket{E-}$. This of course suggests a general tendency to find particularly
strongly entangled polarons in dots with symmetry related degeneracies.

\begin{table}[h]
  \begin{center}
    \begin{tabular}{|c|c|c|c|}
      \hline
        \ \ State\ \   & Entanglement & Phonon Number & Relax. Factor \\
      \hline
        $\ket{A_1,1}$  & 0.023 & 0.042 & 0.001 \\
      \hline
        $\ket{A_1,3}$  & 0.007 & 0.963 & 0.007 \\
      \hline
        $\ket{A_1,6}$  & 0.523 & 0.995 & 0.520 \\
      \hline
        $\ket{A_1,2}$  & 0.203 & 0.920 & 0.186 \\
      \hline
        $\ket{A_1,4}$  & 0.222 & 0.084 & 0.019 \\
      \hline
        $\ket{A_1,5}$  & 0.517 & 0.996 & 0.515 \\
      \hline
        $\ket{E\pm,1}$ & 0.256 & 0.136 & 0.035 \\
      \hline
        $\ket{E\pm,2}$ & 0.245 & 0.902 & 0.221 \\
      \hline
        $\ket{E\pm,3}$ & 0.229 & 0.962 & 0.220 \\
      \hline
    \end{tabular}
    \caption{Entanglement, 1-phonon probability and heuristic ``relaxativity measure''
    of the Quantum Dot Polarons (QDPs) in the pyramidal dot with $h = 10mn$.}
    \label{tab entanglement}
  \end{center}
\end{table}

Finally, we address a possible connection between entanglement and phonon-mediated decoherence and relaxation. A simple
model of such decoherence and relaxation would account for a weak bulk interaction between coupled LO-phonons and
uncoupled LO-phonons or between coupled LO-phonons and LA-phonons. Such weak interactions maybe treated perturbatively
and typically result in a finite lifetime for QDPs, which would otherwise be everlasting. One may expect that the
lifetime depends on the entanglement between electronic and phononic coordinates, since entanglement indicates strong
quantum correlations that could effectively translate phonon-phonon interactions to electron state hoppings. From this
picture, we expect that the lifetime also scales with the weight of the phonon component. Hence, we heuristically
propose a ``relaxativity measure'' for QDPs, defined as the product of the entanglement and the average phonon number
\begin{eqnarray}
  Rel\big(\ket{\Gamma\,j,m}\big) &\equiv &\bra{\Gamma\,j,m}
  \sum_{\lambda\lambda'} (\Lambda^{-1})_{\lambda'\lambda}B_\lambda^\dag B_{\lambda'}
   \ket{\Gamma\,j,m} \nonumber \\
  & & \; \times Ent\big(\ket{\Gamma\,j,m}\big)
\end{eqnarray}
In the present one-phonon model this measure varies between 0 (everlasting) and 1 (short coherence time, say $\sim 1
ps$). At thermal equilibrium, the dot state is represented by a density matrix exhibiting a high probability of states
with a low relaxativity measure and vise versa. This measure does not really measure the relaxation since relaxation
also implies other factors like resonances and population of final states, this is why we speak of ''relaxativity''. It
has the status of a rough heuristic guess, since it is not the result of a proper relaxation model describing
realistically phonon-phonon interactions and in particular neglects any dependence on the particular geometry.

\subsection{Dot size variation}\label{dot size variation}

\begin{figure}[h]
  \includegraphics{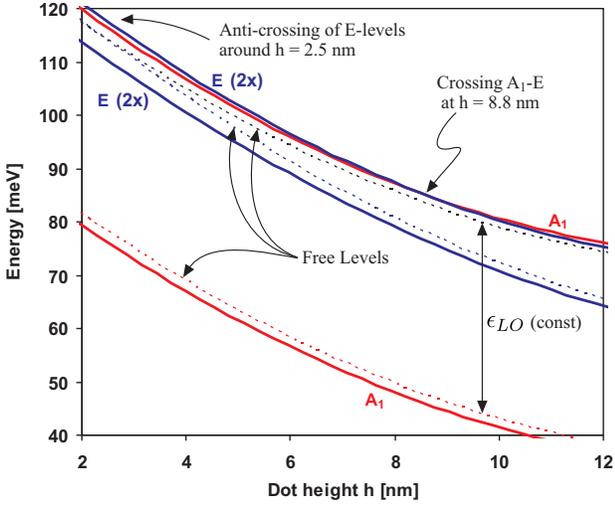}
  \caption{(Color online) Spectrum in function of the dot height $h$. (dashed lines) free electronic energies obtained by
  diagonalizing $\hfree$, (solid lines) polaron energies obtained by diagonalizing $\hqdp$.}
  \label{fig 09}
\end{figure}

We shall now discuss the variation of the polaron spectrum as a function of a
varying dot height $h$. Figure~\ref{fig 09} shows the varying spectrum of
$\hfree$ (free levels) and the spectrum of $\hqdp$ (quantum dot polarons) in
the restricted energy band $[40\ meV,\ 120\ meV]$ (quadratically extrapolated
from explicit computations of the dot heights 10 nm, 7.5 nm and 5 nm). To gain
clarity and to remain consistent with the disappearance of certain free levels
for smaller dots, we have restricted the graph to the two lowest levels of the
two subspaces $\Hrelonesymstrong{A_1}{1}$ and $\Hrelonesym{E+}$. The latter is
of course degenerate with $\Hrelonesym{E-}$ and the respective levels are twice
degenerate.

There are three relevant free energies, the electronic ground level with
symmetry $A_1$ (red dashed line), the first excited level with symmetry $E$
(blue dashed line), and the electronic ground level combined with one phonon
(black dashed line). The latter two undergo a crossing in the vicinity of the
dot height $h=2.5nm$. By virtue of the resulting resonance, the two $E$-like
polaron levels (blue solid lines) exhibit maximal energy shifts around
$h=2.5nm$ ($\Delta\epsilon\approx 5\ meV$) giving rise to a level
anti-crossing. On the other hand, the decreasing resonance for increasing dot
height, leads to a true crossing between the first excited $E$-like polaron level
(upper blue solid line) and the first excited $A_1$-like polaron level (upper
red solid line). Such a true crossing is consistent with the strict analytical
decoupling stemming from group theoretical arguments (\textit{i.e.} different irreps).


\section{Insights on the low energy scheme in QDs}\label{generalQD}

We shall now expand the results to a very general class of QDs, including
pyramidal, spherical, cubic or even cylindrical ones. For all these systems we
uncover an analogous low energy spectrum, clear connections between polarons
and free levels, symmetry properties and qualitative dot size dependencies.

\begin{figure}[h]
  \includegraphics{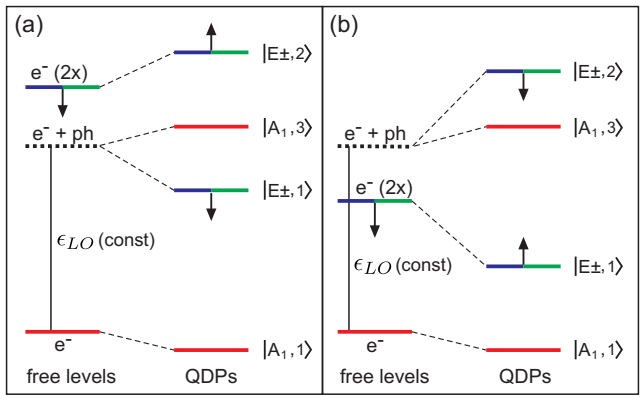}
  \caption{(Color online) Fundamental level structure for QDs with $C_{nv}$-like symmetry with $n\geq 3$. (a) electronic
  energy spacing exceeds phonon energy $\epsilon_{LO}$, typically smaller dots. (b) electronic energy
  spacing smaller than $\epsilon_{LO}$, typically larger dots. The arrows show the level changes
  with increasing dot size.}
  \label{fig 10}
\end{figure}

Explicitly, we consider all dots with a non-degenerate electronic ground level and a twice degenerate first electronic
excitation. These dots include the special but predominant class of dots with $C_{nv}$ symmetry with $n\geq 3$.
Qualitatively, they yield a low-energy polaron spectrum consisting of two shifted electron levels and a splitted
electron+phonon level, see Fig.~\ref{fig 10}. Group theory reveals three independent substructures (red, green, blue),
where two are mutually degenerate (green/blue). This structure can be derived from the natural basis (\ref{sym natural
basis}), or may be obtained from the spectrum studied above (Fig.~\ref{fig 07}) by suppressing all QDPs with higher
energies or associated with the third electron level.

The fine dashed lines in Fig.~\ref{fig 10} link each QDP level with the free level, from which it would arise, if one
could continuously turn on the Fr\"{o}hlich interaction. These connections are important for understanding the QDP
spectrum, as levels within the same representation (here the levels with the same color) generally repel each other
under the interaction. The relative position of the first excited polaron in the invariant representation (here
$\ket{A_1,3}$) depends on whether the free electronic energy spacing is larger or smaller than the constant phonon
energy $\epsilon_{LO}$, see Fig.~\ref{fig 10}a and Fig.~\ref{fig 10}b. In case (a), the state $\ket{A_1,3}$ can fall
between the second electron level (green/blue) and the electron+phonon level (dashed level). In (b), the same state
lies always above the free electron+phonon level. One can generally pass from situation (a) to (b) by a dot size
increase rendering the electronic energy spacing smaller than $\epsilon_{LO}$.

Further, the links between free levels and QDP levels allow to predict the variation of QDP levels with varying dot
size. In general, the shifts become larger as one approaches the resonance, which is the transition between case (a)
and (b). The varying spacing between the upper two free levels leads to a changing shift of the two degenerate QDP
levels (green/blue). These changes are represented by the vertical arrows for increasing dot size. The dependance gets
reversed when passing from case (a) to (b), due to the anti-crossing at the resonance. The shift of the two symmetrical
polaron levels (red) is dot size independent, because of their symmetry-decoupling from the moving free level
(green/blue) and the constant phonon energy $\epsilon_{LO}$.

These findings are very generic. For example they agree with the results of Verzelen \textit{et al.}~\cite{Verzelen 00}
for the case of cylindrical dots (height/radius=12/18). Case (a) is obtained for radii $<13$ nm, whilst case (b)
correspond to radii $>13$ nm. Following our discussion, it is straightforward to understand that in their case
$\ket{1\pm}$ lies below $\ket{S1}$, that $\ket{2\pm}$ lies above $\ket{P\pm 0}$ and that $\ket{\tilde{S}1}$ can only
lie below $\ket{P\pm 0}$ for radius $<13$ nm. We also see that the shifts of $\ket{\tilde{S}0}$ and $\ket{\tilde{S}1}$
do not depend on the dot size due to symmetry decoupling and the constant spacing between $\ket{S0}$ and $\ket{S1}$.

\section{Summary}\label{conclusion}

In this work we uncovered the substantial advantage of the direct use of non-orthogonal creation and annihilation
operators to treat polarons in quantum structures. Starting from a general viewpoint, we fully reformulated the polaron
problem in terms of those operators that naturally appear in the interaction Hamiltonian and generate the phonons
relevant for individual transitions between electronic eigenstates. We also provided a complementary basis for all
non-coupling phonons, which play a sensitive role in relaxation processes mediated by phonon-phonon interactions. Even
though one might a priori be skeptic with the use of non-orthogonal objects, this approach proved mathematically
elegant and fruitful for physical insights. In particular, we found a nested structure in the electron-phonon coupling,
which allowed us to identify a non-trivial rule to truncate the Hilbert space in the case of a finite number of
phonons. This feature was consistently applied to a general QD structure in a one-electron/one-phonon model, and lead
to a novel polaron basis, baptized the ``natural basis''. The latter constitutes an efficient tool for computation and
detailed classification of quantum dot polarons (QDPs). Beyond the case of general quantum dots, we also investigated
degenerate and symmetrical quantum dots using the appropriate mathematical instruments, namely group theory. This
revealed additional simplifications, degeneracies and subclasses of QDPs.

As a realistic application we computed the low-energy QDPs of recently manufactured pyramidal QDs with
$C_{3v}$-symmetry. To this end an adaptive irregular discretization of the lattice mode space was developed, which we
used to compute the Fr\"{o}hlich matrix elements. The generalized eigenvalue problem stemming from the direct use of
non-orthogonal basis vectors was directly fed into efficient matrix diagonalization software. In this way, the
requirement for computational resources was remarkably decreased. The numerical results explicitly revealed the
spectral structure predicted from the natural basis. 3D-visualizations of the stationary polaronic dot states gave
insight in the localization of both electronic and phononic components and showed the different symmetry properties.
Dot size dependent spectral investigations uncovered level crossings and anti-crossings, which were consistent with the
corresponding symmetry properties. Further, we could prove the existence of strong coupling regimes for each symmetry
representation through explicit comparison with second order perturbation theory. Yet, there was undoubtable numerical
evidence for the presence of very weakly coupled subspaces within the strong coupling regimes. This led us to the
concept of ``weakly coupled strong coupling regimes''. Using the natural basis such subspaces could be understood in
terms of specifically different overlaps between electronic wave functions and non-orthogonal vibrational modes. We
used Bennett's entanglement measure to quantify the coupling between electronic and phononic coordinates $-$ an idea
that finally lead us to a heuristic ``relaxativity measure''. In the end, we discussed the low-energy spectrum of an
important class of symmetric QDs (including spherical, cubic and cylindrical dots), and showed qualitative predictions
of the level structure and dot size dependance, valid as much for the general case as for our specific $C_{3v}$
pyramidal QD.

We thank Pawe{\l} Machnikowski and our referees for their thorough and inspiring suggestions. Further, we acknowledge
partial financial support from the Swiss NF project No.\,200020-109523.

\section{Appendix}\label{appendix}

\subsection{Derivation of the coefficients $c_{\q\q'}$}\label{derivation 1}
In section \ref{1st}, the coefficients $c_{\q\q'}$ were defined as
\begin{equation}\label{xx}
  B_\q^\dag \equiv \sum_{\q'}{c_{\q\q'}{b_{\q'}}^\dag}
\end{equation}
such that $B_\q^\dag\ket{0}= (\openone-\mathcal{P})\,b_\q^\dag\ket{0}$ with $\mathcal{P}$ being the orthogonal
projector onto $vect\{B_\lambda^\dag\ket{0}\}$. By substitution we find
\begin{equation}\label{aa}
  \sum_{\q'}{c_{\q\q'}{b_{\q'}}^\dag}\ket{0}=(\openone-\mathcal{P})\,b_\q^\dag\ket{0}
\end{equation}

The projector $\mathcal{P}$ satisfies
\begin{equation}\label{d}
  \mathcal{P}B^\dag_\lambda\ket{0}=B^\dag_\lambda\ket{0}
\end{equation}
and can be decomposed as
\begin{equation}\label{e}
  \mathcal{P}=\sum_{\lambda\lambda'}p_{\lambda\lambda'}B^\dag_\lambda\ket{0}\bra{0}B_{\lambda'}
\end{equation}
with complex coefficients $p_{\lambda\lambda'}$. Substituting (\ref{e}) in (\ref{d}) and using the commutation
relations (\ref{main commutator}) yields the unique solution $p_{\lambda\lambda'}=(\Lambda^{-1})_{\lambda\lambda'}$.
Hence,
\begin{equation}\label{g}
\begin{split}
  \mathcal{P}&=\sum_{\lambda\lambda'}(\Lambda^{-1})_{\lambda\lambda'}B^\dag_\lambda\ket{0}\bra{0}B_{\lambda'}\\
             &=\sum_{\q\q'}\sum_{\lambda\lambda'}(\Lambda^{-1})_{\lambda\lambda'}L^\ast_{\lambda\q}L_{\lambda'\q'}
             b^\dag_\q\ket{0}\bra{0}b_{\q'}\\
\end{split}
\end{equation}

Substituting (\ref{g}) into (\ref{aa}) finally exhibits the unique solution
\begin{equation}
  c_{\q\q'} = \delta_{\q\q'}-\sum_{\lambda\lambda'}(\Lambda^{-1})_{\lambda\lambda'}L^\ast_{\lambda\q'}L_{\lambda'\q}
\end{equation}

\subsection{Demonstration of equation (\ref{demo2})}\label{derivation 2}
We want to find the $p$-phonon part $\HQSPxkph{p}$ of a subspace $\mathcal{S}$, defined as
\begin{equation}
  \mathcal{S}\equiv vect\Big\{e^{-i\hqsp t}\ket{\varphi} : \forall\,t,\ \forall\,\ket{\varphi}\in\HQSPkph{p-1}\Big\}
\end{equation}
In the following, we implicitly assume that $\ket{\varphi}$ goes over all states of $\HQSPkph{p-1}$ (or, equivalently,
over a basis of $\HQSPkph{p-1}$). In the expansion of the exponential, the functions $1,t,t^2,t^3,\ldots$ are linearly
independent. Thus,
\begin{equation}
  \mathcal{S}=vect\Big\{(\hqsp)^k\ket{\varphi}\,\ \forall\ k=0,1,\ldots\Big\}
\end{equation}
We then replace $\hqsp$ by $\hfree+\hfr_++\hfr_-$, where $\hfr_+$ is the phonon creating term of $\hfr$ and $\hfr_-$ is
the phonon annihilating term ($\hfr=\hfr_++\hfr_-$),
\begin{equation}
  \mathcal{S}=vect\Big\{(\hfree+\hfr_++\hfr_-)^k\ket{\varphi}\,\forall\,k=0,\ldots\Big\}\
\end{equation}

As we are interested in the $p$-phonon subspace of $S$, the terms $(\hfree+\hfr_++\hfr_-)^k$ can be significantly
simplified by retaining only the operator products increasing the phonon number by one unit. These are the products,
which contain exactly one more $\hfr_+$ than $\hfr_-$. Further, we want to respect the assumed truncation of the phonon
Fock space to at most $p$ phonons, that is imposing $B^\dag_\lambda\ket{\varphi}=0\ \forall\
\ket{\varphi}\in\HQSPkph{p}$ (see section \ref{coupling structure}). Explicitly, we need to remove all products
involving intermediate $(p+1)$-phonon states (e.g.\ $\hfr_-{\hfr_+}^2\ket{\varphi}$, which involves the state
${\hfr_+}^2\ket{\varphi}$). Applying these rules, the terms $(\hfree+\hfr_++\hfr_-)^k\ket{\varphi}$ reduce to
\begin{eqnarray*}
    k=1 &:& \hfr_+\ket{\varphi} \\
    k=2 &:& (\hfree\hfr_++\hfr_+\hfree)\ket{\varphi} \\
    k=3 &:& ({\hfree}^2\hfr_++\hfree\hfr_+\hfree+\hfr_+{\hfree}^2 \\
        & & +{\hfr_+}^2\hfr_-+\hfr_+\hfr_-\hfr_+)\ket{\varphi} \\
    etc. &&
\end{eqnarray*}
Since these vectors are used to span a collective subset, we can merely clean the list by creating new superpositions.
Explicitly, we walk down the list from $k=1,2,\ldots$ and subtract all the parts that are manifestly covered by smaller
$k$ already. For $k=2$, for example, we can subtract $\hfr_+\hfree\ket{\varphi}$ from
$(\hfree\hfr_++\hfr_+\hfree)\ket{\varphi}$, since $\hfr_+\hfree\ket{\varphi}=\hfr_+\ket{\varphi'}$ with
$\ket{\varphi'}\in\HQSPkph{p-1}$ is already spanned by the vectors associated with $k=1$. Hence the additional vectors
from $k=2$ can be reduced to $\hfree\hfr_+\ket{\varphi}$. ($vect\{\hfr_+\ket{\varphi}\}$ and
$vect\{\hfree\hfr_+\ket{\varphi}\}$ are not necessarily linearly independent, but together they certainly span the same
subspace as all the vectors in the list associated with $k=1$ and $k=2$.) We can then apply the same procedure to $k=3$
and find that all terms but ${\hfree}^2\hfr_+\ket{\varphi}$ are manifestly spanned by the vectors of $k=1$ and $k=2$.
One quickly realizes that proceeding in the same way, subsequently produces all the terms
${\hfree}^3\hfr_+\ket{\varphi}$, ${\hfree}^4\hfr_+\ket{\varphi}$, etc. Hence,
\begin{equation}
  \HQDPxkph{p}=vect\Big\{(\hfree)^k\hfr_+\ket{\varphi}\ \forall\ k=0,1,\ldots\Big\}
\end{equation}
Using again the property that $1,t,t^2,t^3,\ldots$ are linearly independent functions of $t$, we finally find
\begin{equation}
  \HQDPxkph{p}=vect\Big\{e^{-i\hfree t}\hfr_+\ket{\varphi}\ \forall\ t\Big\}
\end{equation}
which concludes the demonstration.

\subsection{Expression of  $\hfr$  for an irregular q-space discretization}\label{froehlich}

In the Fr\"{o}hlich matrix elements (\ref{froehlich matrix element}), the
quantization volume $V$ (direct space) dictates the underlying $\q$-space
discretization, such that each $\q$ occupies a volume of $\Omega=(2\pi)^3/V$.
This can be seen by taking $V$ as a cubic volume with periodic boundary
conditions, for which the Fr\"{o}hlich interactions was originally derived. If
we use an \emph{irregular} space discretization with varying cell sizes
$\Omega(\q)$, the constant quantization volume $V$ must consequently be replace
by a function,
\begin{equation}
  V\rightarrow V(\q)=\frac{8\pi^3}{\Omega(\q)}
\end{equation}
In the present case, $\Omega(\q)$ was taken as the Wigner-Seitz volume around
the point $\q$ in a given irregular reciprocal space discretization.

\subsection{$C_{3v}$-Symmetrized Phonon Basis}\label{c3v phonons}
We consider the symmetry group $C_{3v}$ with its six group elements $g=I$ (identity), $g=C^+_3,C^-_3$ (positive and
negative $2\pi/3$-rotation), $g=\sigma_1,\sigma_2,\sigma_3$ (plane symmetries). If $\ket{\q}$ denotes the one-phonon
state associated with the plane wave mode $\xi_\q(\x)$, symmetrized one-phonon states $\ket{\Gamma,j,\q}$ can be
obtained by
\begin{equation}
  \ket{\Gamma,j,\q}\equiv\alpha\mathcal{P}_{\Gamma,j}\ket{\q}=\sum_{g\in C_{3v}} c_{\,\Gamma j}(g)\ket{\mathcal{R}(g)\q}
  \label{def symmetrized phonon states}
\end{equation}
where $\mathcal{R}(g)$ is the symmetry operation associated with the group element $g$. $\mathcal{P}_{\Gamma,j}$ is the
projector on the subspace associated with the irrep $\Gamma$ and the partner function j. $\alpha$ is a normalization
factor defined up to a phase factor by the normalization relations
\begin{equation}
  \braket{\Gamma,j,\q}{\Gamma,j,\q}=1\quad\forall\ \Gamma,j,\q\
  \label{normalization relation}
\end{equation}

By projecting the $n_{modes}$ basis states $\ket{\q}$ on the four subspaces associated with $A_1$, $A_2$, $E+$ and
$E-$, one obtains an overcomplete set of $4n_{modes}$ states, which necessarily obeys $3n_{modes}$ relations of linear
dependence. Those relations can be identified with the symmetry transformation relations,
\begin{equation}
  A_1: \ket{A_1,\mathcal{R}(g)\q} = \ket{A_1,\q}\quad\forall g\in C_{3v}\label{A1}
\end{equation}
\begin{equation}
  A_2: \ket{A_2,\mathcal{R}(g)\q} = \left\{
  \begin{array}{l}
    +\ket{A_2,\q}\quad g=I,C^+_3,C^-_3 \\
    -\ket{A_2,\q}\quad g=\sigma_1,\sigma_2,\sigma_3
  \end{array}
  \right. \label{A2}
\end{equation}
\begin{equation}
 E: \begin{array}{l}
    \ket{E\pm,\mathcal{R}(\sigma_1)\q} = \pm\ket{E\pm,\q} \\
    \ket{E\pm,\q}+\ket{E\pm,\mathcal{R}(C^+_3)\q}+\ket{E\pm,\mathcal{R}(C^-_3)\q} = 0
   \end{array} \label{E}
\end{equation}

\begin{figure}[h]
  \includegraphics{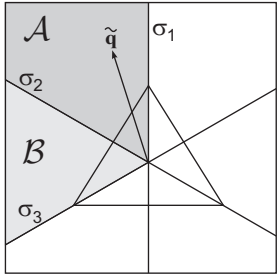}
  \caption{Partition of the set of available wave vectors $\q$. The whole set can be obtained by applying
  symmetry operations of the $C_{3v}$ group on the subset $\mathcal{A}$.}
  \label{fig 11}
\end{figure}

For the irrep $A_1$ the five non-trivial relations of (\ref{A1}) for a given $\q$ allow to restrict the plane wave set
$\{\q\}$ to the sixth marked $\mathcal{A}$ in Fig.~\ref{fig 11}. The set $\{\ket{A_1,\qt}, \qt\in\mathcal{A}\}$ is
orthonormal. An analog reasoning applies to the irrep $A_2$ based on the five non-trivial relations of (\ref{A2}). For
the irrep $E$, (\ref{E}) yields two non-trivial relations for a given vector $\q$ and a given partner function $j=\pm$.
Hence the set $\{\q\}$ may be restricted to a third of its elements, represented by $\mathcal{A}\cup\mathcal{B}$ in
Fig.~\ref{fig 11}. Any two states $\ket{E,j,\qt}$ and $\ket{E,j,\mathcal{R}(\sigma_2)\qt}$ with $\qt\in\mathcal{A}$
(and hence $\mathcal{R}(\sigma_2)\qt\in\mathcal{B}$) are non-orthogonal. In order to achieve orthogonality and to use
\emph{one fixed vector set for all irreps}, we introduce the states
\begin{equation}
  \label{new E states}
  \begin{split}
  \ket{E\pm,\qt,\chi=1} & \equiv \ket{E\pm,\qt}\mp\ket{E\pm,\sigma_2\qt} \\
  \ket{E\pm,\qt,\chi=2} & \equiv \ket{E\pm,\qt}\pm\ket{E\pm,\sigma_2\qt} \\
  \end{split}
\end{equation}
where $\qt\in\mathcal{A}$ and hence $\mathcal{R}(\sigma_2)\qt\in\mathcal{B}$. This definition completes the
construction of the one phonon part of the phonon basis (\ref{phonon basis}). The new index $\chi=1,2$ permits the
restriction of plane wave vectors to the sixth $\mathcal{A}$ and has the following physical interpretation: All
$E$-states with $\chi=1$ involve \emph{one} plane wave amplitude, whereas the $E$-states with $\chi=2$ mix \emph{two}
different amplitudes (see Fig.~\ref{fig 06}c).

The canonical transformation relating the symmetrized one-phonon
basis to the plane wave basis $\{\ket{\qt}\}$ results from the
definitions (\ref{def symmetrized phonon states}) and (\ref{new E
states}) and the normalization relation (\ref{normalization
relation}). We find,
\begin{equation}\label{basis transformation}
\left(%
\begin{array}{l}
  \ket{A_1,\qt} \\
  \ket{A_2,\qt}  \\
  \ket{E,+,\qt,1}  \\
  \ket{E,+,\qt,2}  \\
  \ket{E,-,\qt,1}  \\
  \ket{E,-,\qt,2} \\
\end{array}%
\right)=U
\left(%
\begin{array}{l}
  \ket{\qt}  \\
  \ket{\mathcal{R}(\sigma_2)\qt}  \\
  \ket{\mathcal{R}(C^+_3)\qt}  \\
  \ket{\mathcal{R}(\sigma_3)\qt}  \\
  \ket{\mathcal{R}(C^-_3)\qt}  \\
  \ket{\mathcal{R}(\sigma_1)\qt} \\
\end{array}%
\right)\quad\forall\ \qt\in\mathcal{A}
\end{equation}
with the unitary transformation matrix
\begin{equation}
U=\left(%
\begin{array}{cccccc}
  \frac{1}{\sqrt{6}} & \frac{1}{\sqrt{6}} & \frac{1}{\sqrt{6}} & \frac{1}{\sqrt{6}} & \frac{1}{\sqrt{6}} & \frac{1}{\sqrt{6}}  \\
  \frac{1}{\sqrt{6}} & -\frac{1}{\sqrt{6}} & \frac{1}{\sqrt{6}} & -\frac{1}{\sqrt{6}} & \frac{1}{\sqrt{6}} & -\frac{1}{\sqrt{6}}  \\
  \frac{1}{2} & -\frac{1}{2} & 0 & 0 & -\frac{1}{2} & \frac{1}{2}  \\
  \frac{1}{\sqrt{12}} & \frac{1}{\sqrt{12}} & -\frac{1}{\sqrt{3}} & -\frac{1}{\sqrt{3}} & \frac{1}{\sqrt{12}} & \frac{1}{\sqrt{12}}  \\
  \frac{1}{2} & \frac{1}{2} & 0 & 0 & -\frac{1}{2} & -\frac{1}{2}  \\
  \frac{1}{\sqrt{12}} & -\frac{1}{\sqrt{12}} & -\frac{1}{\sqrt{3}} & \frac{1}{\sqrt{3}} & \frac{1}{\sqrt{12}} & -\frac{1}{\sqrt{12}} \\
\end{array}
\right)
\end{equation}

\subsection{$C_{3v}$-Symmetrized Tensor Product Basis}\label{c3v tensor product basis}

Based on the symmetrized electron basis (\ref{electron basis}) and
the symmetrized phonon basis (\ref{phonon basis}), we shall
construct symmetrized product states. The zero-phonon state
$\ket{0}$ belonging to the irrep $A_1$, the
symmetrized product states involving zero phonons readily write,
\begin{equation}
  \begin{split}
  \ket{A_1;A_1,\alpha_e;0} & \equiv \ket{A_1,\alpha_e} \otimes \ket{0} \\
  \ket{E\pm;E;0}           & \equiv \ket{E\pm} \otimes \ket{0} \\
  \end{split}
\end{equation}
where $n=1,2$ is the electronic energy index inside $A_1$. Semicolons separate intrinsic polaron, electron and phonon
indices. As for the symmetrized product states involving one phonon, one uses Clebsch-Gordan coefficients,
\begin{equation*}
  \begin{array}{lll}
    \ket{A_1;A_1,\alpha_e;A_1,\qt}      & \equiv & \ket{A_1,\alpha_e} \otimes \ket{A_1,\qt}  \\
    \ket{A_1;E;E,\qt,\chi}        & \equiv & \frac{1}{\sqrt{2}}\Big(\ket{E+}\otimes\ket{E+,\qt,\chi} \\
                                 && + \ket{E-}\otimes\ket{E-,\qt,\chi}\Big)  \\
    \ket{A_2;A_1,\alpha_e;A_2,\qt}      & \equiv & \ket{A_1,\alpha_e} \otimes \ket{A_2,\qt} \\
    \ket{A_2;E;E,\qt,\chi}        & \equiv & \frac{1}{\sqrt{2}}\Big(\ket{E+}\otimes\ket{E-,\qt,\chi} \\
                                 && - \ket{E-}\otimes\ket{E+,\qt,\chi}\Big)  \\
    \ket{E\pm;A_1,\alpha_e;E,\qt,\chi}   & \equiv & \ket{A_1,\alpha_e}\otimes\ket{E\pm,\qt,\chi}  \\
    \ket{E\pm;E;A_1,\qt}         & \equiv & \ket{E\pm}\otimes\ket{A_1,\qt}  \\
    \ket{E\pm;E;A_2,\qt}         & \equiv & \ket{E\mp}\otimes\ket{A_2,\qt}  \\
    \ket{E\pm;E;E,\qt,\chi}       & \equiv & \frac{1}{\sqrt{2}}\Big(\ket{E+}\otimes\ket{E\pm,\qt,\chi} \\
                                 & & \mp \ket{E-}\otimes\ket{E\mp,\qt,\chi}\Big)
  \end{array}
\end{equation*}
where $\chi=1,2$ is the additional phonon index used for $E$-like one-phonon states.

\end{document}